\documentclass[11pt]{article}
\usepackage{fullpage}
\usepackage{times}
\usepackage[margin=1in]{geometry}
\usepackage{amsmath,amsfonts,amssymb,amsthm}
\usepackage{tikz}
\usepackage{graphicx}
\usepackage{enumerate}
\usepackage{etoolbox}
\usetikzlibrary{decorations, patterns}
\newcommand{\E}{\ensuremath{\mathbb{E}}}

\newcommand{\ORC}{\mathsf{2D\text{-}ORC}}
\newcommand{\TDORC}{\mathsf{3D\text{-}ORC}}
\newcommand{\ANC}{\mathsf{4ANC}}
\newcommand{\coMA}{\mathsf{coMA^{cc}}}
\newcommand{\MA}{\mathsf{MA^{cc}}}

\newcommand{\cA}{\mathcal{A}}
\newcommand{\cB}{\mathcal{B}}
\newcommand{\cC}{\mathcal{C}}
\newcommand{\cE}{\mathcal{E}}
\newcommand{\cI}{\mathcal{I}}
\newcommand{\cP}{\mathcal{P}}
\newcommand{\cG}{\mathcal{G}}
\newcommand{\cH}{\mathcal{H}}
\newcommand{\cM}{\mathcal{M}}
\newcommand{\cN}{\mathcal{N}}
\newcommand{\cX}{\mathcal{X}}
\newcommand{\cY}{\mathcal{Y}}
\newcommand{\cS}{\mathcal{S}}
\newcommand{\cQ}{\mathcal{Q}}
\newcommand{\cU}{\mathcal{U}}
\newcommand{\cT}{\mathcal{T}}
\newcommand{\cR}{\mathcal{R}}
\newcommand{\cL}{\mathcal{L}}
\newcommand{\cF}{\mathcal{F}}
\newcommand{\cZ}{\mathcal{Z}}
\newcommand{\cD}{\mathcal{D}}

\newcommand{\cO}{\mathcal{O}}
\newcommand{\cW}{\mathcal{W}}

\newcommand{\tP}{\widetilde{P}}
\newcommand{\tp}{\widetilde{p}}

\newcommand{\eps}{\epsilon}
\newcommand{\Rtau}{R_{\tau}}
\newcommand{\Xtau}{X_{\tau}}
\newcommand{\Ytau}{Y_{\tau}}
\newcommand{\Rpi}{R_{\pi}}
\newcommand{\Xpi}{X_{\pi}}
\newcommand{\Ypi}{Y_{\pi}}
\newcommand{\mm}{m_{\textrm{meg}}}
\newcommand{\MM}{M_{\textrm{meg}}}
\newcommand{\mml}{m_{\textrm{mer}}}
\newcommand{\hmml}{\hat{m}_{\textrm{mer}}}
\newcommand{\MML}{M_{\textrm{mer}}}
\newcommand{\Sx}{S^{\hmml}_x}

\newcommand{\cm}{column-monochromatic}
\newcommand{\typ}{evenly-spreading}
\newcommand{\iso}{isolated}
\newcommand{\ws}{well-separated}
\newcommand{\dec}{\mathrm{DEC}}
\newcommand{\tq}{\ensuremath{\widetilde{Q}}}

\newcommand{\SIMD}{\textrm{SIM}_D}

\newcommand{\suc}[3]{\mathsf{Suc}^{#1}_{#2}\left(#3\right)}

\newcommand{\algwidth}{0.97\textwidth}

\newcommand{\mnote}[1]{ \marginpar{\tiny\bf
            \begin{minipage}[t]{0.5in}
              \raggedright #1
           \end{minipage}}}
           
\def\mainfile{}

\newtoggle{focs}
\ifx\focs\undefined
	\togglefalse{focs}
\else
	\toggletrue{focs}
\fi

\begin{document}

\newtheorem{definition}{Definition}
\newtheorem{theorem}{Theorem}
\newtheorem{lemma}{Lemma}
\newtheorem{proposition}{Proposition}
\newtheorem{conjecture}{Conjecture}
\newtheorem{claim}{Claim}
\newtheorem{corollary}{Corollary}
\newtheorem*{remark}{Remark}
\newtheoremstyle{restate}{}{}{\itshape}{}{\bfseries}{~(restate).}{.5em}{\thmnote{#3}}
\theoremstyle{restate}
\newtheorem*{restate}{}

\title{
Amortized Dynamic Cell-Probe Lower Bounds from Four-Party Communication}

\author{Omri Weinstein\thanks{Department of Computer Science, New York University. Supported by a Simons Society Junior Fellowship.} \and 
Huacheng Yu\thanks{Department of Computer Science, Stanford University. Supported by NSF CCF-1212372.}}
\date{}

\maketitle 

\thispagestyle{empty}
\addtocounter{page}{-1}

\begin{abstract}
This paper develops a new technique for proving amortized, randomized cell-probe lower bounds on dynamic 
data structure problems. We introduce a new randomized nondeterministic four-party communication model 
that enables ``accelerated", error-preserving simulations of dynamic data structures.   

We use this technique to prove an $\Omega(n\left(\log n/\log\log n\right)^2)$ cell-probe lower bound for the 
dynamic 2D  weighted orthogonal range counting problem ($\ORC$) with $n/\mathrm{poly}\log n$ updates 
and $n$ queries, that holds even for data structures with $\exp(-\tilde{\Omega}(n))$ success probability. 
This result not only proves the highest amortized lower bound to date, but is also tight in the strongest possible 
sense, as a matching upper bound can be obtained by a deterministic data structure with worst-case operational 
time. This is the first demonstration of a ``sharp threshold" phenomenon for dynamic data structures.  

Our broader motivation is that cell-probe lower bounds for exponentially small success facilitate \emph{reductions 
from dynamic to static} data structures. As a proof-of-concept, we show that a slightly strengthened version of our 
lower bound would imply an $\Omega((\log n /\log\log n)^2)$ lower bound for the \emph{static} $\TDORC$ problem 
with $O(n\log^{O(1)}n)$ space. Such result would give a near quadratic improvement over the highest known static 
cell-probe lower bound, and break the long standing $\Omega(\log n)$ barrier for static data structures.  

\end{abstract}


\newpage

\section{Introduction}
\ifx\mainfile\undefined
\documentclass[11pt]{article}
\usepackage{fullpage}
\usepackage{times}
\usepackage[margin=1in]{geometry}
\usepackage{amsmath,amsfonts,amssymb,amsthm}
\usepackage{graphicx}
\usepackage{color}
\newcommand{\E}{\ensuremath{\mathbb{E}}}

\newcommand{\ORC}{\mathsf{2D\text{-}ORC}}
\newcommand{\TDORC}{\mathsf{3D\text{-}ORC}}
\newcommand{\ANC}{\mathsf{4ANC}}
\newcommand{\R}{\mathbb{R}}
\newcommand{\coMA}{\mathsf{coMA^{cc}}}
\newcommand{\MA}{\mathsf{MA^{cc}}}

\newcommand{\algwidth}{0.97\textwidth}

\newcommand{\cP}{\mathcal{P}}
\newcommand{\cW}{\mathcal{W}}

\newcommand{\mnote}[1]{ \marginpar{\tiny\bf
            \begin{minipage}[t]{0.5in}
              \raggedright #1
           \end{minipage}}}

\begin{document}

\newtheorem{definition}{Definition}
\newtheorem{theorem}{Theorem}
\newtheorem{corollary}{Corollary}
\newtheorem{lemma}{Lemma}
\newtheorem{proposition}{Proposition}
\newtheorem{conjecture}{Conjecture}
\newtheorem{claim}{Claim}
\newtheorem*{remark}{Remark}

\section{Introduction}
\fi
\def\introfile{}

Understanding 
the limitations of data structures in the cell-probe model \cite{Yao81} is one of the holy grails of theoretical computer science, 
primarily since this model imposes very weak implementation constraints and hence captures 
essentially any imaginable data structure.  Unfortunately, this abstraction makes it notoriously difficult to obtain 
lower bounds on the operational time of data structures, 
in spite of nearly four decades of active research. 
For dynamic data structures, where a sequence of $n$ database operations (interleaved updates and queries) is to be correctly maintained, 
the highest \emph{amortized} cell-probe lower bound to date is $\Omega(\log n)$ per operation, i.e.,  $\Omega(n\log n)$ 
for a sequence of $\Theta(n)$ operations (P\v{a}tra\c{s}cu and Demaine~\cite{PD06}\footnote{Notably, this bound holds 
only for high-probability data structures which succeed on solving all queries with probability $1-n^{-\Omega(1)}$.}).  
The breakthrough work of Larsen \cite{Larsen12a} brought a near-quadratic improvement for \emph{worst-case} number of probes per operation. 
Larsen gave an $\Omega((\log n/\log\log n)^2 )$ query time lower bound for the dynamic \emph{weighted orthogonal range
counting} problem in two-dimensional space ($\ORC$), which holds for any data structure with at most polylogarithmic update time. 
In this fundamental problem, the data structure needs to maintain a set of weighted points in the two-dimensional plane, and support the following operations: 
\begin{itemize}
\item
\texttt{update($r$, $c$, $w$)}: insert a point at $(r, c)$ with weight $w$,
\item
\texttt{query($r$, $c$)}: the sum of weights of points dominated by $(r,c)$,\footnote{A point $(r',c')$ is dominated by $(r,c)$ if $r'\leq r$ and $c'\leq c$.}
\end{itemize}
where $r, c, w\in [n]$.\footnote{$[n]$ stands for the set of integers $\{1,2,\ldots,n\}$.}
Larsen's aforementioned bound is tight when $O(\log^{2+\epsilon} n)$ update time is allowed, as there is a deterministic data structure that solves 
the problem using $O(\delta\log^2 n)$ probes per update and $O((\log_{\delta} n)^2 )$ probes per query in the worst-case for any $\delta>1$.  However, it is often the case 
that amortization can reduce the average cell-probe complexity (a notable example is the Union Find problem \cite{Tarjan75, Blum85}), 
especially when randomization is permitted and the data structure is allowed to err with constant probability \emph{per query}.

Indeed, one particular shortcoming of all known dynamic  data structure lower bounds is that they are not robust to error: 
All previous lower bounds only apply to deterministic, Las-Vegas or at most constant-error randomized data structures (e.g., \cite{FS89, PD06,PT11, Yu15h}). 
A more robust question, which we motivate below, is to study the rate of decay of success probability in answering all (or most) of 
the queries, as a function of the allocated resources (in our context, the total number of probes). 
The distinction above is similar in spirit to the difference between ``direct sum" theorems in complexity theory 
(e.g., \cite{FederKNN95, KarchmerKN95, PT06, BBCR10}) which assert a lower bound on the number of resources 
required for solving multiple instances of a given problem with \emph{constant overall success}, and ``direct product" theorems such as the celebrated 
parallel repetition theorem \cite{Raz98} and Yao's XOR lemma \cite{Yao82}, which 
further asserts an exponential decay in the success probability if insufficient resources are provided. 
Beyond unravelling the nature of ``parallel computation", one of the primary motivations of direct product theorems is black-box hardness amplification 
(see e.g., \cite{DinurS14} and references therein). 
In the context of the cell-probe model, we argue that such theorems open a new path for proving 
both dynamic and static data structure lower bounds, via \emph{reductions} (more on this below). 

Despite the long history of direct product theorems in complexity theory (see e.g. \cite{JainPP12} and references therein),  we are not aware of any 
such result in the cell-probe model.\footnote{It is noteworthy that, unlike direct product theorems in other computational models such as 
two-prover games \cite{Raz98}, circuit complexity \cite{Yao82} and interactive models 
and proof systems \cite{JainPP12, BRWY13}, the dynamic cell-probe model is closer to the setting of \emph{sequential repetition}, since the model is 
online: The data structure needs to provide answers to one query before it receives the next. This feature potentially makes the problem harder than 
``parallel repetition"  (where all problem instances appear in a ``batch").} 
Indeed, a crucial assumption which direct sum and product theorems rely on  
is the premise that  all  copies of the   problem are  \emph{independent} of each other. 
Alas,  in the dynamic data structure model, all 
$q$  queries $Q_1 , Q_2,\ldots  ,Q_q$ 
are essentially with respect to the \emph{same} (or 
slightly modified) database $X$! 
Due to this (asymmetric) correlation, one should not expect generic (``black-box") 
direct product theorems for arbitrary dynamic data structure problems, 
and such a surprising result may only be true due to the specific structure of the underlying problem.  
The main result of this paper asserts that the $\ORC$
problem exhibits such interesting structure, leading to the following strong amortized lower bound: 

\def\thmtwodStatement{For any integer $n$, $1\leq c< o(\log n/\log\log n)$, and any (randomized) data structure $D$ in the cell-probe model with word-size $\Theta(\log n)$, 
there is a sequence of $n/\log^c n$ updates and $n-n/\log^c n$ queries 
for the $\ORC$ problem, for which the probability (over the randomness of $D$) that
\begin{itemize}
	\item
		$D$ probes $o(n\left(\log n/c\log\log n\right)^2)$ cells in total, and
	\item
		$D$ is correct on all $n-n/\log^c n$ queries
\end{itemize}
is at most $2^{-n/\log^{c+O(1)} n}$.}
\begin{theorem}[Amortized Lower Bound for $\ORC$]\label{thm2d}
\thmtwodStatement
\end{theorem}

Theorem~\ref{thm2d} not only provides a near quadratic improvement over the previous highest amortized cell-probe lower bound,
but it is also tight in the strongest possible sense, as it exhibits  
 a ``sharp threshold" phenomenon for $\ORC$: 
while $O(n\left(\log n/c\log\log n\right)^2)$ probes are sufficient to solve the problem \emph{deterministically}, Theorem \ref{thm2d} asserts that  
any dynamic data structure that spends 
$\ll\! n\left(\log n/c\log\log n\right)^2$ probes will have success probability which is hardly any better than the trivial success probability of 
\emph{randomly guessing the answers to all queries}!
To best of our knowledge, this is the first result of its kind in the cell-probe model.  

We note that it is possible to modify our proof of Theorem~\ref{thm2d} so that the lower bound holds even if the second condition is relaxed to 
``$D$ is correct on $99\%$ of the $n-n/\log^c n$ queries''. In many realistic dynamic scenarios, where the data structure is executed as 
a sub-procedure that supports a long sequence of (possibly multi-user) applications (e.g., routing, navigation and other network computations), this relaxed 
error criteria is more suitable and much less restrictive than requiring the data structure to succeed on all queries with an overall probability of $99\%$. 
Nevertheless, this ``Chernoff-type'' variant of Theorem~\ref{thm2d} rules out efficient dynamic data structures for $\ORC$ 
even under this substantially more modest and realistic requirement. 

The broader agenda we suggest and promote in this paper is that 
proving dynamic cell-probe lower bounds for data structures with exponentially small success probability 
facilitates \emph{reductions from dynamic to static}  
data structure problems. The general outline of such reduction is as follows: 
suppose we can show that any (randomized) dynamic data structure for a problem $\cP$ 
that has at least $\exp(-u)$ success probability in answering 
a sequence of queries with $u$ updates, must probe at least $t$ cells. 
We would like to argue that a \emph{static} data structure $D$ with a too-good query time 
for some static problem related to $\cP$, must use a lot of space. Indeed, $D$ 
can be used to solve the \emph{dynamic} problem $\cP$ with $> \exp(-u)$ probability, 
simply by \emph{guessing} all $u$ updates, preprocessing them and storing in the memory in advance, 
which in turn would imply that $D$ must use a at least $\Omega(t)$ memory cells. 
Since in the dynamic problem $\cP$, updates and queries are \emph{interleaved}, answering the $i$th query $Q_i$ requires knowing precisely 
those updates \emph{preceding} $Q_i$ in the sequence. This means that $D$ must guess (and store) not only the updates themselves, but also 
the \emph{time} (i.e., order) at which they occurred. One way to incorporate this extra information is to add an extra ``time coordinate" to each query and update 
of the problem $\cP$, which results in a slightly augmented (static) problem $\cP^+$. In general, $\cP^+$ might not correspond 
to any natural data structure problem, however, when $\cP = \ORC$, this extra ``time coordinate" can  be embedded as a \emph{third} dimension 
of the (weighted, two-dimensional) points, in which case the augmented static problem $\cP^+$ corresponds to nothing else but the 
\emph{three-dimensional} weighted orthogonal range counting problem ($\TDORC$).  As a proof-of-concept of the approach above, 
we show that 
if the bound in Theorem~\ref{thm2d} can be slightly strengthened so that it holds for even smaller success probability 
(by a polylogarithmic factor in the exponent), then the following breakthrough result would follow for static $\TDORC$: 

\begin{proposition}[
From dynamic $\ORC$ to static $\TDORC$]\label{prop3d_reduction} \label{prop_3dorc}
Suppose the probability in Theorem~\ref{thm2d} can be 
further reduced to 
$n^{-3n/\log^c n}=2^{-3n/\log^{c-1} n}$. Then any (zero-error) static data structure for $\TDORC$ that uses $n\log^{O(1)} n$ space, 
requires $\Omega\left((\log n/\log\log n)^2\right)$ query time. 
\end{proposition}

In contrast, the best static lower bound to date for orthogonal range counting (in any dimension) is only $\Omega(\log n/\log\log n)$, even 
for \emph{linear}-space data structures. In fact, no $\omega(\log m)$ lower bound is known for any static data structure problem, where $m$ 
is the range of the queries (e.g., $m=n^2$ for $\ORC$ and $m=n^3$ for $\TDORC$). 
So while the slightly stronger premise of Proposition~\ref{prop_3dorc} 
appears to be non-trivial to prove (see the discussion in Appendix~\ref{discussion}), 
if this approach can be realized, it would yield a near-quadratic improvement in static cell-probe lower bounds. 
The formal proof of Proposition~\ref{prop_3dorc} can be found in Appendix~\ref{sec_append_prop_3dorc}.   

We remark that the aforementioned reduction is merely an example, while other reductions (e.g., between different dynamic problems) may be possible via similar outline. 
More generally, if one can show that, conditioned on some (low probability) event $\cW$, a solution to problem $A$ produces a solution to problem $B$, 
then ruling out efficient data structures for problem $B$ with $\approx p(\cW)$ success, would yield a cell-probe lower bound for $A$ as well.  

In the remaining subsections of this introduction, we provide a brief outline of 
the new techniques we develop en-route to proving Theorem~\ref{thm2d}, and how they overcome limitations of 
previous techniques used in the dynamic cell-probe model. 

\subsection{Related work and previous techniques}
\ifx\introfile\undefined
\documentclass[11pt]{article}
\usepackage{fullpage}
\usepackage{times}
\usepackage{amsmath,amsfonts,amssymb,amsthm}
\usepackage{graphicx}
\usepackage{color}
\newcommand{\E}{\ensuremath{\mathbb{E}}}

\newcommand{\ORC}{\mathsf{2D\text{-}ORC}}
\newcommand{\TDORC}{\mathsf{3D\text{-}ORC}}
\newcommand{\ANC}{\mathsf{4ANC}}
\newcommand{\R}{\mathbb{R}}
\newcommand{\coMA}{\text{co-}\mathbf{MA^{cc}}}
\newcommand{\MA}{\mathbf{MA^{cc}}}

\newcommand{\algwidth}{0.97\textwidth}

\newcommand{\cP}{\mathcal{P}}

\newcommand{\mnote}[1]{ \marginpar{\tiny\bf
            \begin{minipage}[t]{0.5in}
              \raggedright #1
           \end{minipage}}}

\begin{document}

\fi
\newcommand{\bU}{\mathbf{U}}
 
Several techniques have been developed along the years for proving lower bounds 
on the cell-probe complexity of dynamic data structure problems.
This line of work was aimed not just at proving lower bounds for a broader class of problems, 
but also at facilitating higher lower bounds for stronger and more realistic data structures (e.g., randomized, amortized).
In most problems and applications, it is natural to assume that the description of an operation can fit in $O(1)$ words,
and the most natural assumption on the word-size of the cell-probe model is $w=\Theta(\log n)$. 
In this regime, Fredman and Saks~\cite{FS89} first introduced the \emph{chronogram} method, 
and used it to prove an $\Omega(\log n/\log \log n)$ lower bound for 
the 0-1 partial sum problem. 
This lower bound stood as a record 
for 15 years, until P\v{a}tra\c{s}cu and Demaine~\cite{PD04a} 
introduced the \emph{information transfer tree} technique 
which led to a tight 
$\Omega(\log n)$ lower bound for general partial sum, improving the highest lower bound by a factor of $\log\log n$.
About a decade later, Larsen~\cite{Larsen12a} showed how to combine the chronogram method with the \emph{cell-sampling} technique 
(which was used for proving \emph{static} data structure lower bounds~\cite{PTW10}), and proved an $\Omega((\log n/\log\log n)^2)$ 
worst-case lower bound
for $\ORC$. 
In the natural regime, this is also the highest lower bound proved for \emph{any} explicit problem hitherto. 
In the remainder of this subsection, we outline Larsen's approach and the challenges in extending his techniques to the type of dynamic lower bounds we seek. 

We remark that other lower bounds have been proved for the regime where $w=\omega(\log n)$. 
In particular, P\v{a}tra\c{s}cu \cite{Patrascu07} proved a matching $\Omega((\log n/\log\log n)^2)$ lower bound for the $\ORC$ problem, 
but only when both the weights of points and word-size are $\log^{2+\epsilon} n$ bits long (we elaborate on the connection between this result and our techniques in Section \ref{subset_intro_anc}).

\paragraph*{Larsen's approach.} 
To prove the aforementioned $\Omega((\log n/\log\log n)^2)$ lower bound for $\ORC$, one considers a sequence of $n$ random updates. 
The idea is to show that after these $n$ updates have been performed, a random query must probe many cells. 
More specifically, the $n$ updates are partitioned into $\Theta(\log n/\log\log n)$ \emph{epochs}: $\ldots, \bU_i,\ldots,\bU_2, \bU_1$, 
where the $i$-th epoch $\bU_i$ consists of $\beta^i$ updates for $\beta=\mathrm{poly}\log n$. The goal is to show that in expectation, 
a random query must read $\Omega(\log n/\log\log n)$ memory cells that are written during epoch $i$, but never overwritten later. 
Let us restrict the attention to epoch $i$ and assume that all updates in other epochs are fixed arbitrarily (i.e., only $\bU_i$ is random). 
Let $S_i$ denote the set of cells whose last update occurred in epoch $i$. 
Indeed, any cell that is written \emph{before} epoch $i$ cannot contain any information about $\bU_i$,
while the construction guarantees that there are few cells written \emph{after} epoch $i$, due to the 
exponential decay in the lengths of epochs.
 Thus, one concludes that ``most" of the information the data structure learns about 
$\bU_i$ comes from cell-probes to $S_i$.
Then the basic idea is to sample a subset $C_i\subseteq S_i$ of a \emph{fixed} size. 
Then for each query, the fewer cells in $S_i$ the data structure probes when answering it, the more likely that all of them will belong to the random subset $C_i$. 
Thus, if a random query probes too few cells in $S_i$ (in expectation), 
there will be too many queries that can be answered without probing any cell in $S_i\setminus C_i$. 
One then argues that the answers to these queries reveal too much information about $\bU_i$, even more than they should: all cells in $C_i$ can contain at most $|C_i|\cdot w$ bits of information. 
This yields a lower bound on the number of cells a random query must probe in $S_i$, and implies a query time lower bound.
\\

The above approach relies on the fact that update time has a \emph{worst-case} upper bound. Indeed, 
the statement that ``very few cells are probed after $\bU_i$''  may no longer hold when we only have an amortized guarantee on the update 
time, because the data structure could spend a long time on epoch $\bU_1$ (say). 
In fact, if we allow amortization for updates, the above sequence of operations is no longer hard, since the data structure 
can simply record each update until the last one, and then spend $O(n\log n)$ time to construct a static $\ORC$ data structure that operates in  
$O(\log n)$ query time. 
Over the $n$ updates, it only spends $O(\log n)$ time per update ``on average''. Obviously, this is not a good dynamic data structure in general, 
because it is not even in a ready-to-query state until the very end. 

To prove an amortized lower bound, it is therefore necessary to \emph{interleave} queries and updates as in~\cite{PD04a,PD06,PT11,Yu15h}. We observe that a variation of Larsen's 
approach can be adapted to prove a zero-error-data-structure version of Lemma~\ref{lem_lower_probe}. Combining this version of the lemma with our proof of Theorem~\ref{thm2d} would yield an alternate proof of our amortized lower bound for zero-error data structures. However, it seems highly non-trivial to 
generalize this proof so that it applies to data structures with exponentially small success probability. 
Roughly speaking, on the one hand, the cell-sampling technique appears to be inapplicable for simultaneous analysis of multiple queries 
as it only applies to a fixed memory state, 
whereas in our setup different queries are performed on different memory states. 
On the other hand, the ``direct product"  lower bound we seek requires 
analyzing the \emph{conditional} success probability (and performance) of a given query, conditioned on success in previous queries. 
Conditioning on this event may leak a lot of information about previous updates, making the proof much more subtle and hard to analyze
(note that this was not an issue for zero-error data structures!).


%
%
\ifx\introfile\undefined
\end{document}
\fi

\subsubsection{Communication-based techniques for dynamic lower bounds}

One of the successful approaches for proving dynamic data structure lower bounds relies on reductions from the 
\emph{communication complexity}  
model, where the general idea is to partition the operation sequence between Alice and Bob and the communication task is to answer all 
queries in Bob's operation interval. To prove a (meaningful) lower bound on the number of probes required by any data structure operating 
over the operation sequence, one needs to show that Alice and Bob can efficiently simulate any data structure for the dynamic problem, 
so that a data structure with too few probes induces a too-good-to-be-true protocol for the communication game (the problem 
then boils down to proving a communication lower bound for the communication problem, which is often easier to analyze). 
For the simulation to be fast, Bob needs to be able to efficiently obtain 
the (memory contents of)  ``relevant" cells probed in his interval that were updated 
during Alice's operation interval.

Indeed, the choice of the communication model is crucial for the simulation argument: If the communication model is too weak, the simulation 
would be ``too slow" 
for proving a strong (or even meaningful) cell-probe lower bound; On the other hand, if the communication model is too strong, proving a high 
lower bound on the amount of communication may be extremely difficult or even impossible (as we elaborate below). 
P\v{a}tra\c{s}cu \cite{Patrascu07} used the standard two-party randomized communication model to prove an $\Omega((\log n/\log\log n)^2 )$ 
cell-probe lower bound on $\ORC$, but only for (somewhat unnatural) weight-size and word-size $w=\Theta(\log^{2+\epsilon}n)$. 
This caveat stems from his simulation being ``too slow": 
P\v{a}tra\c{s}cu's simulation argument requires Alice to send a very long message 
(a \emph{Bloom Filter} of length $\approx\! \log^2 n$ bits per operation) in order for Bob to 
figure out the aforementioned set of ``relevant" cells, hence for the simulation to produce a non-trivial communication lower bound, Alice's input had better dominate the 
latter communication step, which is precisely why points are chosen to have $(\log^{2+\epsilon}n)$-bit weights.

P\v{a}tra\c{s}cu and Thorup \cite{PT11} somewhat remedied this by introducing simulation in the two-party \emph{nondeterministic} communication 
model, in which a know-all ``prover" (Merlin) can help the players reduce their communication by providing some (untrusted) \emph{advice} 
which requires verification. While this technique can be used to speed up the simulation process (leading to new dynamic lower 
bound for several data structure problems), it turns out to be still too slow for the type of lower bounds we seek (in particular for range-counting problems). 
But even more importantly, the nondeterministic reduction of \cite{PT11} does not readily extend beyond \emph{zero-error} (Las-Vegas) data structures. 
Indeed, when the data structure and hence the simulating protocol are allowed to err, the simulation above naturally leads to \emph{randomized nondeterministic} 
communication models such as $\MA \cap \coMA$. Proving strong lower bounds on such powerful models is a notoriously hard open problem 
(see e.g., \cite{Klauck11}), and in our case may even be impossible (indeed, the $\ORC$ problem is related to computation of inner-products over finite fields,  which 
in turn admits a surprisingly efficient ($\tilde{O}(\sqrt{n})$ bit) $\mathsf{MA}$-protocol \cite{AW08}).


\subsection{Our techniques and the $\ANC$ communication model} \label{subset_intro_anc}
We introduce a new randomized nondeterministic communication model (which we hence term $\ANC$) that solves both problems above, namely, 
it enables faster (error-preserving\footnote{When seeking lower bound for data structures with tiny (exponentially small) success, such reductions 
must not introduce any (non-negligible) error, or else the soundness of the reduction is doomed to fail.}) simulations of randomized data structures, 
yet in some aspect is much weaker than $\MA$, and hence amenable to substantial lower bounds.   
To enable a faster simulation (than \cite{PT11,Yu15h}), our model includes \emph{two} provers (hence four parties in total) who are communicating only with Bob: 
The first prover (Merlin) is \emph{trusted} but has limited ``communication budget", while the second prover (Megan) is \emph{untrusted}, yet has \emph{unlimited}  
``communication budget". More precisely, the model requires Alice and Bob's computation to be correct \emph{only when Merlin is ``honest"}, 
but charges for each bit sent by Merlin (hence the model is only meaningful for computing two-party functions with large range, as Merlin can 
always send the final answer and the players would be done).  In contrast, the model doesn't charge for Megan's message length, but requires 
Bob to \emph{verify} that her message is  correct (with probability 1!).  
Intuitively, the model allows Bob to receive some short ``seed" of his choice (sent by Merlin), in such way that this ``seed" 
can be used to extract much more information (a longer message sent by Megan) in a verifiable (i.e., consistent) way.

We show that this model can indeed help the players ``speed up" their simulation:
Merlin can send a succinct message (the ``seed", which in the data structure simulation would correspond to some ``succinct encoding" of 
memory addresses of the relevant  (intersecting) cells probed by the data structure in both Alice and Bob's operation intervals),  
after which Megan can afford to send a \emph{significantly longer}  message (which is supposed to be the actual memory addresses 
of the aforementioned cells). 
Alice can then send Bob all the relevant \emph{content} of these cells (using communication proportional to the \emph{number of 
``relevant" cells probed} by the data structure (times $w$)).  
If both Merlin and Megan are honest, Bob has all the necessary information to answer his queries. Otherwise, if Megan is cheating 
(by sending inconsistent addresses with Merlin's seed), we argue that Bob can detect this during his simulation
given all the information he received from Merlin and Alice, yielding a fast and admissible $\ANC$ protocol.  

For our setting of the parameters, this simulation saves a $\mathrm{poly}\log(n)$ factor of communication for Alice, and $\approx\! \log n$ factor of 
communication for Bob, compared to the standard nondeterministic simulations of \cite{PT11,Yu15h}.  
We stress that this speed-up is essential to prove the cell-probe lower bound we seek on $\ORC$, 
and is most likely to be important in future applications. 

To solve the second problem, namely, to limit the power of the model (so that it is amenable to substantial lower bounds), 
we impose two important constraints on the non-deterministic advice 
of Merlin and Megan: Firstly, the provers can only talk to Bob, hence the model is \emph{asymmetric} ;
Secondly and most importantly, we require that the provers' advice are \emph{unambiguous}, i.e., Merlin's (honest) message is uniquely determined 
by some pre-specified function of the player's inputs, and similarly, for each message sent by Merlin (whether he tells the truth or not),  
Megan's (honest) message is uniquely specified by the players' inputs and Merlin's message. 
These restrictions are tailored for data structure simulations, since for any (deterministic) data structure $D$, the aforementioned set of ``relevant" 
memory cells probed by $D$ is indeed a deterministic function of the operation sequence. 

We show that these two features imply a generic structural fact about $\ANC$ protocols, namely, that low-communication protocols 
with \emph{any nontrivial success} probability in this model
induce large biased (i.e., ``semi-monochromatic") rectangles in the underlying communication matrix (see Lemma \ref{lem_monochromatic_rect}). 
Intuitively, this follows from the uniqueness property of the model, which in turn implies that for any fixed messages sent by Merlin, 
the resulting protocol induces a partition of the input matrix into \emph{disjoint} biased rectangles. 
In contrast, we remark that rectangles induced by $\MA$ protocols may \emph{overlap} (as there may be multiple transcripts that 
correspond to the same input $(x,y)$), which is part of why proving strong lower bounds on $\MA$ is so difficult. 

Therefore, ruling out efficient randomized communication protocols (and hence a too-good-to-be-true data structure) for a given communication 
problem boils down to ruling out large biased rectangles of the corresponding communication matrix. 
Since we wish to prove a lower bound for protocols with tiny (exponentially small) 
success for $\ORC$, we must rule out rectangles with exponentially small ``bias".  
This ``direct-product" type result for $\ORC$ in the $\ANC$ model (Lemma \ref{lem_2DORC_cc_lb}) 
is one of the main steps of the proof of Theorem \ref{thm2d}.

\ifx\mainfile\undefined
\bibliographystyle{alpha}
\bibliography{refs}

\end{document}

\fi

\subsection{Organization}

\iftoggle{focs}{
Due to space constraints, the next 5 pages of this abstract only contain a high-level proof of our results, where most proofs are deferred to the appendix. 
We begin by formally defining the $\ANC$ model in Section \ref{sec_anc}. 
We then prove that $\ANC$ protocols can efficiently simulate dynamic data structures (Section \ref{subsect_sim_general}), and on the other hand, 
that efficient $\ANC$ protocols induce large biased rectangles of the underlying communication matrix (Section \ref{subsect_cm_rect}). 
In Section \ref{sec_ds_lb} we state our main technical lemma  which rules out such rectangles (even with exponentially small bias) for $\ORC$ (Lemma \ref{lem_2DORC_cc_lb}), 
and finally tie the pieces together to conclude the proof of Theorem \ref{thm2d}.
}{
We begin by formally defining the $\ANC$ model in Section \ref{sec_anc}. 
We then prove that $\ANC$ protocols can efficiently simulate dynamic data structures (Section \ref{subsect_sim_general}), and on the other hand, 
that efficient $\ANC$ protocols induce large biased rectangles of the underlying communication matrix (Section \ref{subsect_cm_rect}). 
In Section \ref{sec_ds_lb} we prove our main technical lemma  which rules out such rectangles (even with exponentially small bias) for $\ORC$ (Lemma \ref{lem_2DORC_cc_lb}), 
and finally tie the pieces together to conclude the proof of Theorem \ref{thm2d}.
}

\iftoggle{focs}{}{

\section{Preliminaries}\label{sect_pre}

\subsection{The Cell-Probe Model}

A dynamic data structure in the cell-probe model consists of an array of memory cells, each of which can store $w$ bits. 
Each memory cell is identified by a $w$-bit address, so the set of possible addresses is $[2^w]$. It is natural to assume that 
each cell has enough space to address (index) all update operations performed on it, hence we assume that 
$w=\Omega(\log n)$ when analyzing a sequence of $n$ operations. 

Upon an update operation, the data structure can perform read and write operations to its memory so as to reflect the update, 
by \emph{probing} a subset of memory cells. This subset may be an arbitrary function of the update and the content of the memory 
cells previously probed during this process. The update time of a data structure is the number of probes made when processing an 
update (this complexity measure can be measured in worst-case or in an amortized sense). 
Similarly,  upon a query operation, the data structure can perform a sequence of probes to read a subset of the memory cells in order 
to answer the query. Once again, this subset may by an arbitrary (adaptive) function of the query and previous cells probed during the 
processing of the query. The query time of a data structure is the number of probes made when processing a query. 

\subsection{Communication Complexity}

In the classical two-party communication complexity model \cite{Yao79}, two players (Alice and Bob) receive inputs 
$x\in \cX$ and $y\in \cY$ respectively (possibly from some joint distribution $(x,y)\sim \mu$), and need to collaborate to solve some joint function $f: \cX\times\cY\to \cZ$ of their inputs. 
To do so, they engage in an interactive communication \emph{protocol} $\pi$. In round $i$, one player (which must be specified by the protocol) 
sends the other player a message $m_i$.
In a \emph{deterministic} protocol, $m_i$ is a function  of the previous transcript $m_{<i}$ and the input of the player ($x$ if Alice is the 
speaker in round $i$ and $y$ if Bob is the speaker). In a (public-coin) \emph{randomized} protocol, messages may further depend on 
a public random string $r$ which is observed by both players (when the protocol or inputs are randomized, we sometimes use $\Pi$ to denote the 
(random variable) corresponding to the transcript of $\pi$).  
The \emph{communication cost} of $\pi$ is the (worst-case) number of bits transmitted 
in the protocol in any execution of $\pi$ (over $x,y,r$). 

The \emph{distributional communication complexity}  of $f$ with respect to input distribution $\mu$  
and success $\delta$ is the minimum communication cost of a (deterministic) protocol which correctly solves $f(x,y)$ with probability $\geq \delta$ over $\mu$.
The \emph{randomized communication complexity}  of $f$
is the minimum communication cost of a protocol which correctly solves $f(x,y)$ \emph{for all inputs} $x,y$,  with probability $\geq \delta$ over the public 
randomness $r$ of the protocol. The two measures are related via Yao's minimax theorem \cite{Yao77, Yao79}.  
The following basic definitions and properties of communication protocols are well known (see \cite{BookKN97} for a more thorough exposition).

\begin{definition}[Communication matrix]
The communication matrix $M(f)$ of a two-party function $f: \cX\times\cY\to \cZ$, is the matrix indexed by rows and columns $x\in \cX,y\in \cY$, whose entries 
are $M(f)_{x,y} = f(x,y)$. 
\end{definition}

\begin{definition}[Combinatorial rectangles and monochromatic rectangles]
A \emph{combinatorial rectangle} (or simply, a rectangle) of $M(f)$ is a subset $R = X\times Y$ of inputs, such that $X\subseteq \cX$, $Y\subseteq \cY$. 
A rectangle $R$ is said to be \emph{monochromatic} if $f$'s value is fixed on all inputs $(x,y)\in R$. 
\end{definition}

The following weaker definition will be more suitable for measuring error in the asymmetric randomized $\ANC$ model we define in this paper:
\begin{definition}[$\alpha$-column-monochromatic rectangles] \label{def_col_mon}
Let $\mu$ be a joint distribution over inputs $(x,y)$. 
A rectangle $R = X\times Y$ of $M(f)$ is said to be \emph{$\alpha$-column-monochromatic} with respect to $\mu$, if for every $y\in Y$,  
at least an $\alpha$-fraction of the entries in column $y$ of $R$ have the same value in $M(f)$, i.e., 
there is some function value $v_y\in \cZ$ such that $\mu((X\times \{y\}) \cap \{f^{-1}(v_y)\}) \geq \alpha\mu(X\times \{y\})$. 
\end{definition}

A basic fact in communication complexity is that a $c$-bit communication protocols that computes $f(x,y)$ 
in the \emph{determinisitc} model, induces a partition of $M(f)$ into at most $2^c$ \emph{monochromatic rectangles}. 
Each rectangle corresponds to a transcript of $\pi$ (i.e., the set of inputs for which this transcript will occur forms a rectangle). 
A similar characterization holds in the \emph{randomized}  or \emph{distributional} models, where rectangles are  ``nearly" 
monochromatic. We will show that protocols in the $\ANC$ model also induce a similar structure on $M(f)$, in terms of biased 
column-monochromatic rectangles (see Lemma \ref{lem_monochromatic_rect}). Hence ruling out large biased column-monochromatic 
rectangles in $M(f)$ can be used to prove communication lower bounds on $f$ in the $\ANC$ model.

}

\section{$\ANC$: A New Four-Party Nondeterministic Communication Model} \label{sec_anc}

We now formally define the $\ANC$ model, which is a randomized, asymmetric, non-deterministic communication model involving 
four players: Alice, Bob, Merlin and Megan. 
Let $\mu$ be a distribution over input pairs $(x,y)\in \mathcal{X}\times \mathcal{Y}$ to Alice and Bob (respectively).
A $\ANC$ protocol $P$ proceeds as follows: In the first stage,  
Alice and Bob use shared randomness to sample a public random string $r$ (of infinite length), which is visible to all four players (Merlin, Megan, Alice and Bob). 
Merlin and Megan observe $(x,y,r)$, and can each send, in turn, a message (``advice") \emph{ to Bob} before the communication 
proceeds in a standard fashion between Alice and Bob. 
As part of the protocol, $P$ specifies, for each input pair and public string $r$, a unique message $\MML(x, y,r)$ that Merlin is supposed to send given that input pair and random string 
(Merlin may not be honest, but we will only require the computation to be correct when he sends the correct message $\MML(x, y,r)$).  
After Merlin sends Bob his message $\mml$ (which may or may not be the ``correct" message $\MML(x, y,r)$), 
it is Megan's turn to send Bob a message. Once again, $P$ specifies (at most)  one message\footnote{Instead of \emph{unique}, $\MM$ can be undefined for 
obviously wrong $\mml$. But $\MM(x, y, \MML(x, y,r),r)$ is always defined.} $\MM(x, y, \mml,r)$ that Megan is supposed to send to Bob, given $x,y,r$ and Merlin's message $\mml$ (as we shall see, the difference between Merlin and Megan's role is that, unlike the case with Merlin's message, the players are \emph{responsible to verify} that Megan's message is indeed correct, i.e., that $\mm = \MM(x, y, \mml,r)$, no matter whether 
$\mml = \MML(x, y,r)$ or not!). In the next stage, Alice and Bob communicate in the standard public-coin communication model, after which Bob decides to 
``proceed" or ``reject" (this is the verification step of Megan's message). Finally, if Bob chooses to proceed, he outputs a value $v$ for $f(x,y)$. 
These stages are formally described in Figure \ref{figure:four_party_prot}. 

\paragraph{Honest Protocol $\tP$.}{Throughout the paper, we denote by $\tP$ the \emph{honest} execution of a $\ANC$ protocol $P$.
More formally, we define $\tp(x,y,r,\tau)$ to be the joint probability distribution of $x,y,r$ and $P$'s transcript $\tau$, when Merlin and Megan send the honest messages (i.e., when $\mml=\MML(x,y,r)$ and $\mm=\MM(x,y,\mml,r)$). 
Note that $\tp$ induces a well defined distribution on transcripts $\tau$, since the transcript of $P$ is completely determined by $(x,y,r)$ in this case. 


\begin{figure}[h!]
\begin{tabular}{|l|}
\hline
\begin{minipage}{\algwidth}
\vspace{1ex}
\begin{center}
\textbf{A 4-party communication protocol $P$}
\end{center}
\vspace{0.5ex}
\end{minipage}\\
\hline
\begin{minipage}[t]{\algwidth}
\vspace{1ex}

\begin{enumerate}
\setcounter{enumi}{-1}
\item
	Alice and Bob generate a public random string $r$, visible to all four players.
\item
	Merlin sends a message $\mml$ to \emph{Bob} ($\mml$ is visible to Megan).
\item
	Megan sends a message $\mm$ to \emph{Bob}.
\item
	Alice and Bob communicate based on their own inputs and $\mml$ and $\mm$ 
	as if they were in the classic communication setting with public randomness.
\item
	Bob decides to \emph{proceed} or \emph{reject}. 
\item
	If Bob chooses to proceed, he outputs a value $v$.
\end{enumerate}

\vspace{1.5ex}
\end{minipage}\\
\hline
\end{tabular}
\caption{A communication protocol $P$ in the $\ANC$ model.}
\label{figure:four_party_prot}
\end{figure}

\begin{definition}[Valid protocols] 
A $\ANC$ protocol $P$ is said to be  \emph{valid} if 
\begin{itemize}
\item
	Bob \emph{proceeds} if and only if $\mm=\MM(x,y,\mml, r)$ \;\;\;\; (with probability 1).
\end{itemize} 
\end{definition}

\begin{definition}[Computation and notation in the $\ANC$ model]\label{def_correctness_anc} 
We say that a $\ANC$ protocol $P$ $\delta$-solves a two-party function 
$f:\mathcal{X}\times\mathcal{Y}\longrightarrow \mathcal{Z}$ with communication cost 
$(c_A, c_B, c_M)$ under input distribution $\mu$ if the following conditions hold.

\begin{enumerate}
\item (Perfect verification of Megan)
	$P$ is a valid protocol.
\item (Communication and correctness)
	With probability at least $\delta$ (over the ``honest" distribution $\widetilde{p}$ and input distribution $\mu$), the honest protocol $\tP$ satisfies that 
	Alice sends no more than $c_A$ bits, Bob sends no more than $c_B$ bits, Merlin sends no more 
	than $c_M$ bits, and Bob outputs the correct value ($v=f(x, y)$).
\end{enumerate}

\noindent For a two-party function $f:\mathcal{X}\times\mathcal{Y}\longrightarrow \mathcal{Z}$ and parameters $c_M,c_A,c_B$, we denote 
by \iftoggle{focs}{$\suc{f}{\mu}{c_A, c_B, c_M}$}{$$\suc{f}{\mu}{c_A, c_B, c_M}$$} the largest probability $\delta$ for which there is a $\ANC$ protocol that $\delta$-solves $f$ under $\mu$ with communication cost $(c_A, c_B, c_M)$.
\end{definition}

\iftoggle{focs}{}{
\begin{remark} \label{rem_anc_model} A few remarks about the model are in order :  
\begin{enumerate}
\item Any function $f:\mathcal{X}\times\mathcal{Y}\longrightarrow \mathcal{Z}$ admits the following three trivial $\ANC$ protocols:
\begin{itemize}
\item
	Alice sends Bob $x$: costs $(\log |\mathcal{X}|,0,0)$.
\item
	Bob sends Alice $y$: costs $(\log |\mathcal{Z}|,\log |\mathcal{Y}|,0)$.
\item
	Merlin sends $f(x, y)$: costs $(0, 0, \log |\mathcal{Z}|)$.
\end{itemize}
\item The players always trust Merlin (since the $\ANC$ model  requires the protocol to be correct only when he is honest).
Nevertheless, Merlin is still allowed to cheat  ($\mml\neq \MML$). 
Even in this case, there is always at most one ``correct'' message Megan should send.  This property will be crucial for the 
characterization of $\ANC$ protocols in terms of monochromatic rectangles (see Subsection \ref{subsect_cm_rect}). 
\item It is important  that Bob is able to verify Megan's message with probability 1, but could be wrong on outputting the function value. 
This corresponds to the requirement that 
the players need to simulate the data structure perfectly, while the data structure itself might succeed with very small probability.
\item Megan is only useful when her advice ($\mm$) is significantly longer than Merlin's advice ($\mml$), as otherwise Merlin might as well 
send Megan's message (and she can remain silent). The benefit here is that the model doesn't charge the protocol for Megan's message 
length (only for verifying it is correct), so when Merlin is honest, Megan helps the players ``speed up" the protocol.
\end{enumerate}	
\end{remark}
}

\subsection{Data structure simulation in the $\ANC$ model}\label{subsect_sim_general}

In this subsection, we show that it is possible to efficiently \emph{simulate} any dynamic data structure on any sequence of operations in the 
$\ANC$ model with no additional error. 
To this end, consider 
a (deterministic) data structure $D$ for some problem $\cP$, and fix a sequence $\cO$ of operations. Let $I_A$ and $I_B$ be two \emph{consecutive} intervals of operations in $\cO$ such that $I_A$ occurs right before $I_B$. Let $P_D(I_A)$ and $P_D(I_B)$ be the set of cells probed by $D$
during $I_A$ and $I_B$ respectively (when $D$ is clear from context, we shall simply write  $P(I_A)$ and $P(I_B)$). Alice is given all operations except for the ones in $I_B$, Bob is given all operations except for the ones in $I_A$. We now describe an $\ANC$ protocol that simulates $D$ on $\cO$ and has the 
same output as $D$ on all queries in $I_B$. 

The naive approach for this simulation is to let Bob simulate the data structure upto the beginning of $I_A$, then skip $I_A$ and continue the simulation in $I_B$. To collect the ``relevant" information on what happened in $I_A$, each time 
$D$ probes a cell that has not been probed in $I_B$ before, Bob asks Alice whether this cell was previously probed in $I_A$, and if it was, he asks Alice to send  the new content of that cell. 
Unfortunately, this approach requires Bob to send $|P(I_B)|\cdot w$ bits and Alice to send $|P(I_B)|+|P(I_A)\cap P(I_B)|\cdot w$ bits. 
However, the players can do much better with Merlin's and Megan's help: 
Merlin reports Bob, upfront, which cells in $P(I_B)$ are probed in $I_A$ in some succinct encoding. 
Given Merlin's succinct message, Megan can send Bob the actual memory addresses of these cells, and the players will be able to easily 
verify these addresses are consistent with the ``seed" sent by Merlin, as the model requires. 
With this information in hand, 
Bob only needs to ask Alice for the contents of \emph{relevant cells} in his simulation, instead of every cell in $P(I_B)$. Moreover, it allows Bob to send this set of cells \emph{in batch}, further reducing his message length.
We turn to describe the formal simulation. 

\paragraph{Protocol $\SIMD$ for Simulating $D$:}

\begin{enumerate}
	\item (Protocol specification of $\MML$.)
		Merlin simulates $D$ upto the end of $I_B$, and generates $P(I_A),P(I_B)$. He sends Bob 
		the sizes $|P(I_A)|$, $|P(I_B)|$ and $|P(I_A)\cap P(I_B)|$.\footnote{Note that although Bob knows all the operations in $I_B$, he still does not 
		know $P(I_B)$, since the operations in $I_A$ are unknown to him, and the data structure can be adaptive.} 
		Then he writes downs the sequence of cells probed during $I_B$ in the chronological order 
		(if a cell is probed more than once, he keeps only the first occurrence). Each cell in the sequence is associated with a bit, indicating whether 
		this cell is also in $P(I_A)$. By definition, this sequence has length $|P(I_B)|$, in which $|P(I_A)\cap P(I_B)|$ cells are associated with a ``1''. 
		Merlin sends Bob the set of indices in the sequence associated with a ``1''.
\iftoggle{focs}{}{In total, Merlin sends 
\[
\begin{aligned}
&O(\log |P(I_A)|+\log |P(I_B)|)+\log {|P(I_B)|\choose |P(I_A)\cap P(I_B)|} \\
\leq\,\,& |P(I_A)\cap P(I_B)|\cdot \log\frac{e|P(I_B)|}{|P(I_A)\cap P(I_B)|}+O\left(\log n\right)
\end{aligned}
\]
		bits. }Note that Merlin's message encodes for each $i$, whether the $i$-th time (during $I_B)$ that $D$ probes a new cell, it was previously probed during $I_A$.
	\item (Protocol specification of $\MM$.)
		Megan simulates $D$ upto the beginning of $I_A$, saves a copy of the memory $M_A$, and continues the simulation upto the beginning of $I_B$, saves a 
		copy of the memory $M_B$. Then she continues to simulate $D$ on $I_B$ \emph{from $\mathbf{M_A}$} with the advice from Merlin. That is, whenever she 
		needs to probe a cell that has not been probed in $I_B$ before, if this is the $i$-th time that this happens and Merlin's message has $i$ encoded in the set, 
		Megan copies the content of the cell from $M_B$ to the current memory, writes down the address of the cell and continues the simulation. Basically Megan 
		simulates $D$ assuming Merlin's claim about which cells probed in $I_B$ are probed in $I_A$ is correct. If there is anything inconsistent during the simulation, 
		$\MM$ is undefined, e.g., $|P(I_B)|$ or $|P(I_A)\cap P(I_B)|$ is different from what Merlin claims, or $D$ breaks during the simulation due to the wrong contents 
		of the memory, etc. As long as Merlin's message is consistent with Megan's simulation, she sends the set of actual memory \emph{addresses} 
		of cells she has written down during the simulation (i.e., the set $P(I_A)\cap P(I_B)$ from Merlin's advice)\iftoggle{focs}{}{ using $|P(I_A)\cap P(I_B)|\cdot w$ bits}. 
	\item (Bob asks the contents of $P(I_A)\cap P(I_B)$.)
		Denote the set of addresses received from Megan by $S$. If $|S| \neq |P(I_A)\cap P(I_B)|$, Bob rejects. Alice and Bob use public randomness to sample a 
		random (hash) function $h:[2^w]\to [|P(I_A)|]$. Bob sends Alice the set of hash-values $h(S)$\iftoggle{focs}{}{ using 
		\[
			\log {|P(I_A)|\choose |P(I_A)\cap P(I_B)|}\leq |P(I_A)\cap P(I_B)|\cdot \log\frac{e|P(I_A)|}{|P(I_A)\cap P(I_B)|}
		\] bits}. 
	\item (Alice replies with the contents.) Alice simulates $D$ and obtains the set $P(I_A)$. For each hash-value $b \in h(S)$, Alice sends Bob both 
	addresses and contents of \emph{all} cells in $P(I_A)$ that 
	are mapped to this value (i.e., of $h^{-1}(b)\cap P(I_A)$).\iftoggle{focs}{}{ Alice sends $4|P(I_A)\cap P(I_B)|\cdot w$ bits in expectation (over the randomness of the hash function).}
	\item (Bob simulates $D$ and verifies Megan.)
		Bob checks whether Alice sends the information about all cells in $S$. If not, he rejects. Otherwise, he simulates the data structure up to the beginning of $I_A$, 
		and then updates all cells in $S$ to the new values. Bob continues the simulation on $I_B$ from this memory state. At last, Bob checks whether the simulation 
		matches Merlin's claim and whether $S$ is exactly the set $P(I_A)\cap P(I_B)$ according to the simulation. If either check fails, he rejects. Otherwise, he proceeds, 
		and generates the output of $D$ on all queries in $I_B$. 
\end{enumerate}

\def\lemSimGeneral{
\begin{lemma}\label{lem_sim_general}
Let $\cO$ be an operation sequence,  $I_A, I_B \subseteq \cO$ be any consecutive operation intervals. Then for any deterministic data structure 
$D$ operating over $\cO$, 
$\SIMD(I_A,I_B)$ is a valid $\ANC$ protocol. Moreover, the honest protocol $\widetilde{\SIMD}$ has precisely the same output as $D$ on 
all queries in $I_B$, 
with Alice sending $$4|P(I_A)\cap P(I_B)|\cdot w$$ bits in expectation, Bob sending at most $$|P(I_A)\cap P(I_B)|\cdot \log\frac{e|P(I_A)|}{|P(I_A)\cap P(I_B)|}$$ bits, 
and Merlin sending at most $$|P(I_A)\cap P(I_B)|\cdot \log\frac{e|P(I_B)|}{|P(I_A)\cap P(I_B)|}+O\left(\log n\right)$$ bits.
\end{lemma}

\begin{proof}
The claimed communication cost of the protocol can be directly verified from steps 1,3 and 4 respectively, so we only need 
to argue about the correctness and validity of the protocol. 
By construction, when both Merlin and Megan are honest, Bob has all the up-to-date information (i.e., latest memory state) 
of the cells $P(I_A)\cap P(I_B)$ probed by $D$ during his operation interval, hence by definition of step 5, $\widetilde{\SIMD}$ 
has the same output as $D$ on queries in $I_B$. It therefore remains to show that $\SIMD$ is a valid $\ANC$ protocol.

To this end, recall that we need to show that for any message $\mml$ sent by Merlin, Bob proceeds iff $\MM(\cO,\mml) = \mm$. 
When Megan is honest (follows the protocol) and sends the set $S$, Bob's simulation of $D$ in step 5 will be exactly the same as Megan's 
in step 2. By definition, $S$ is the exact set of cells that Megan uses the contents from $M_B$ instead of $M_A$. By copying the contents of 
$S$ from Alice's memory state ($M_B$) to Bob's memory state ($M_A$), he recovers Megan's simulation, which is consistent with Merlin's 
message. Thus, Bob will proceed.

When $\MM(\cO,\mml)$ is undefined and Megan follows the protocol and sends the set $S$ she generates in step 2 (but finds inconsistency), 
by the same argument as above, Bob recovers Megan's simulation, thus will find the same inconsistency as Megan does and reject.

The only case left is when Megan chooses to send a different set $S'$ than $S$ (no matter whether $\MM(\cO,\mml)$ is defined).
Let $P_{\mml}(I_B)$ be the set of cells probed during $I_B$ as specified in step 2, given Merlin's advice $\mml$. By definition, 
$S\subseteq P_{\mml}(I_B)$. If $S'\cap P_{\mml}(I_B)=S$ and $S'\neq S$, then by the same argument again, Bob recovers the simulation 
specified in step 2. In the end, he will find that not every cell in $S'$ is probed and reject. Otherwise, consider the symmetric difference 
$(S'\cap P_{\mml}(I_B))\triangle S$. Let $C$ be the first cell in the symmetric difference in the chronological order of probing cells in 
$P_{\mml}(I_B)$, which is the $j$-th new cell probed. Thus, Bob will successfully recover the simulation until he is about to probe cell $C$. 
By definition, $C\notin S'$, if and only if $C\in S$, if and only if $j$ is encoded in $\mml$. Thus, on cell $C$, Bob will find the simulation does 
not match Merlin's claim, and thus reject. 
%
%
\end{proof}
}

\iftoggle{focs}{We defer the analysis of the protocol to Appendix~\ref{app_lem_sim_general}.}{
\lemSimGeneral
}

\subsection{Efficient $\ANC$ protocols induce large biased rectangles}\label{subsect_cm_rect}

Let $P$ be a four-party communication protocol computing $f$ over a  \emph{product} input distribution $\mu=\mu_x\times \mu_y$ in the 
$\ANC$ communication model, with cost $(c_A,c_B,c_M)$ and success probability $\delta$. 
\iftoggle{focs}{The following lemma asserts }{In this section, we are going to prove }that if $P$ 
is efficient (has low communication) and has any ``non-trivial'' accuracy in computing the underlying function $f$, then there must be a large 
biased-column-monochromatic rectangle in the communication matrix of $f$ (see Definition \ref{def_col_mon} for the formal definition). 
We note that a variant of this lemma can be proved for general (non-product) distributions. 
\iftoggle{focs}{Due to space restriction, the proof will be deferred to Appendix~\ref{app_lem_monochromatic_rect}.}{}

\ifx\mainfile\undefined
\documentclass[11pt]{article}
\usepackage{fullpage}
\usepackage{times}
\usepackage{amsmath,amsfonts,amssymb,amsthm}
\usepackage{graphicx}
\usepackage{color}
\newcommand{\E}{\ensuremath{\mathop{\mathbb{E}}}}

\newcommand{\ORC}{\mathsf{2D\text{-}ORC}}
\newcommand{\eps}{\epsilon}
\newcommand{\Rtau}{R_{\tau}}
\newcommand{\Xtau}{X_{\tau}}
\newcommand{\Ytau}{Y_{\tau}}
\newcommand{\Rpi}{R_{\pi}}
\newcommand{\Xpi}{X_{\pi}}
\newcommand{\Ypi}{Y_{\pi}}
\newcommand{\mm}{m_{\textrm{meg}}}
\newcommand{\MM}{M_{\textrm{meg}}}
\newcommand{\mml}{m_{\textrm{mer}}}
\newcommand{\hmml}{\hat{m}_{\textrm{mer}}}
\newcommand{\MML}{M_{\textrm{mer}}}
\newcommand{\Sx}{S^{\hmml}_x}

\newcommand{\cm}{column-monochromatic}

\newcommand{\ANC}{\mathsf{4ANC}}
\providecommand{\Div}[2]{\mathsf{D}\left(\begin{array}{c} #1 \\ \hline  \hline  #2\end{array} \right)} 
\newcommand{\Ex}[2]{\mathop{\mathbb{E}}\displaylimits_{#1}\left
[ #2 \right ]}

\newcommand{\cA}{\mathcal{A}}
\newcommand{\cB}{\mathcal{B}}
\newcommand{\cC}{\mathcal{C}}
\newcommand{\cE}{\mathcal{E}}
\newcommand{\cI}{\mathcal{I}}
\newcommand{\cP}{\mathcal{P}}
\newcommand{\cG}{\mathcal{G}}
\newcommand{\cH}{\mathcal{H}}
\newcommand{\cM}{\mathcal{M}}
\newcommand{\cN}{\mathcal{N}}
\newcommand{\cX}{\mathcal{X}}
\newcommand{\cY}{\mathcal{Y}}
\newcommand{\cS}{\mathcal{S}}
\newcommand{\cQ}{\mathcal{Q}}
\newcommand{\cU}{\mathcal{U}}
\newcommand{\cT}{\mathcal{T}}
\newcommand{\cR}{\mathcal{R}}
\newcommand{\cL}{\mathcal{L}}
\newcommand{\cF}{\mathcal{F}}
\newcommand{\cZ}{\mathcal{Z}}
\newcommand{\cD}{\mathcal{D}}
\newcommand{\cW}{\mathcal{W}}

\newcommand{\tP}{\tilde{P}}
\newcommand{\tp}{\tilde{p}}

\newcommand{\mnote}[1]{ \marginpar{\tiny\bf
            \begin{minipage}[t]{0.5in}
              \raggedright #1
           \end{minipage}}}

\begin{document}

\newtheorem{definition}{Definition}
\newtheorem{theorem}{Theorem}
\newtheorem{lemma}{Lemma}
\newtheorem{proposition}{Proposition}
\newtheorem{conjecture}{Conjecture}
\newtheorem{corollary}{Corollary}
\newtheorem{claim}{Claim}
\newtheorem*{remark}{Remark}

\section{Communication Lower Bound}

\subsection{Efficient $\ANC$ protocols induce large biased rectangles}
\fi


\def\contLemMonochromaticRect{
	$M(f)$ has a rectangle $R=X\times Y$ such that
	\begin{enumerate}
		\item
			$R$ is $\frac{\delta}{2}\cdot 2^{-c_M}$-column-monochromatic; 
		\item
			$\mu_x(X)\geq \frac{\delta}{4}\cdot 2^{-(c_M+c_A+c_B)}$;
		\item
			$\mu_y(Y)\geq \frac{\delta}{4}\cdot 2^{-(c_M+c_B)}$.
	\end{enumerate}
}

\def\contLemMonochromaticRectCompact{
	$M(f)$ has a rectangle $R=X\times Y$ such that: 1) $R$ is $\frac{\delta}{2}\cdot 2^{-c_M}$-column-monochromatic; 2) $
	\mu_x(X)\geq \frac{\delta}{4}\cdot 2^{-(c_M+c_A+c_B)}$; 3) $\mu_y(Y)\geq \frac{\delta}{4}\cdot 2^{-(c_M+c_B)}$.
}

\def\lemMonochromaticRect{
\begin{lemma}[$\ANC$ protocols imply large biased rectangles for product distributions]\label{lem_monochromatic_rect}
\contLemMonochromaticRect
\end{lemma}
}
\def\restateMonochromaticRect{
\begin{restate}[Lemma~\ref{lem_monochromatic_rect}]
\contLemMonochromaticRect
\end{restate}
}
\def\lemMonochromaticRectCompact{
\begin{lemma}[$\ANC$ protocols imply large biased rectangles for product distributions]\label{lem_monochromatic_rect}
\contLemMonochromaticRectCompact
\end{lemma}
}

\def\lemMonochromaticRectFull{

\begin{proof}

Recall 
that $\tP$ denotes the honest execution of the protocol $P$, and by definition of the $\ANC$ model, the probability that 
$\tP$ correctly computes $f(x,y)$ \emph{and} communicates at most $(c_A, c_B, c_M)$ bits respectively, is at least $\delta$.
Since we are working over a fixed input distribution $\mu$, we may fix the public randomness of $P$ to some fixed value ($r=r^*$)  
so that  these conditions continue to hold for the \emph{deterministic} protocol $P_{r^*}$ (over the input distribution $\mu$). 
Define $S$ to be the set of \emph{good} input pairs $(x,y)$ for which $\tP_{r^*}$ correctly computes $f(x,y)$ and the 
the protocol communicates $(c_A, c_B, c_M)$ bits respectively. By definition, $$\mu(S)\geq \delta. $$
For the remainder of the proof, we assume $P$ is deterministic (i.e., we implicitly consider the protocol $P=P_{r^*}$). 
Now consider a transcript $\tau = (\pi,\mm)$ of the deterministic protocol $P$, where $\mm$ denotes  
Megan's message to Bob, and $\pi=(\mml,\pi_A,\pi_B)$ denotes the message from Merlin and the transcript between Alice and Bob. 
For a given message $\mml$ sent by  Merlin, the set of input pairs that will generate the transcript $\tau$ (and for which Bob ``proceeds") form a combinatorial rectangle $\Rtau = \Xtau\times \Ytau$.\footnote{Note that not necessarily every pair in the $\Rtau$ has $\MML=\mml$, as $P$ is not required to verify whether Merlin sends the correct message!} Assuming Bob ``proceeds" in $\tau$, he is supposed to output a value after the communication, which may depend on the transcript and his input $y\in \Ytau$. 
That is, conditioned on Merlin sending $\mml$ (which, once again, may not be the ``honest" message), for each column $y$ of 
the rectangle $\Rtau$ Bob will output the same value. 
Now, recall that for each input $(x, y)$ and $\mml$ there is at most one message $\mm=\MM(x,y,\mml)$ that will make Bob accept
 (i.e., $(x,y,\mml)$ uniquely determine $\mm$, and hence the entire transcript $\tau$). 
Since $P$ is deterministic, this fact implies that if we fix $\mml$, all rectangles $\{\Rtau\}$ are \emph{disjoint} from each other.

Furthermore, since Alice does not observe Megan's message (only Bob does), the set $\Xtau$ does not depend on $\mm$. This means that 
if we fix $\pi$ and vary over all $\mm$ consistent with $\pi$, all rectangles corresponding to resulting transcripts $\tau$ will have 
\emph{the same} $\Xtau$. 
Let $\Rpi=\bigcup_{\tau=(\pi,\mm)}\Rtau$ be the (disjoint) union of these rectangles, which is a rectangle itself, and let $\Xpi\times \Ypi=\Rpi$. 
In this notation, for any $\tau=(\pi,\mm)$ we have that 
$\Xpi=\Xtau$, and $\Ypi=\bigcup_{\tau=(\pi,\mm)}\Ytau$. 
The following claim asserts that every column $y$ of $\Rpi$ will have the same output:

\begin{claim}\label{cl_megan_pi_y}
For each transcript $\pi=(\mml,\pi_A,\pi_B)$ and $y\in \Ypi$, $\MM(x,y,m_{mer})$ is fixed across $x\in \Xpi$. 
\end{claim}

\begin{proof}
Suppose towards contradiction that there is some $y_0\in \Ypi$ and $x_1, x_2 \in \Xpi$, such that 
$$\MM(x_1,y_0,\mml)\neq \MM(x_2,y_0,\mml).$$ 
Now, given $\mm$, Bob's decision whether to ``proceed" or not only depends on his input $y_0$ and the transcript $\pi$
which, by definition, is the same for both inout pairs $(x_1,y_0) , (x_2,y_0)$. 
This means that for at least one of the input pairs, say $(x_1,y_0)$, Bob will ``proceed" even when $\mm = \MM(x_2,y_0,\mml)$, 
contradicting the definition of the $\ANC$ model (proposition $(1)$ in Definition~\ref{def_correctness_anc}). 
\end{proof}

\newcommand{\condi}{\stackrel{\pi=(\mml,\pi_A,\pi_B):}{|\mml|\leq c_M,|\pi_A|\leq c_A,|\pi_B|\leq c_B}}

Indeed, the above claim asserts that Bob's output is only a function of $(\pi,y)$, so let us henceforth denote by $v(\pi,y)$ the output 
of column $y$ of $\Rpi$. Once again, note that for a fixed value of $\mml$, the rectangles $\{\Rpi\}$ are all disjoint. In particular, this fact implies 
\begin{equation}\label{eqn1}    
\sum_{\condi}\mu(R_{\pi})\leq 2^{c_M}.  
\end{equation} 
Now, by definition, every \emph{good} input pair $(x, y)\in S$ is contained in some rectangle $R_{\pi}$, where $\mml=\MML(x, y)$, 
$|\MML(x, y)| \leq c_M$, $|\pi_A|\leq c_A$, $|\pi_B|\leq c_B$ 
and $v(\pi,y)=f(x,y)$ (by definition of $S$). Therefore, we have
\begin{equation}\label{eqn2}
\sum_{\condi}\mu(R_{\pi}\cap S\cap \MML^{-1}(\mml))=\mu(S)\geq \delta.
\end{equation}
\newcommand{\tRpi}{\tilde{R}_{\pi}}
\newcommand{\tYpi}{\tilde{Y}_{\pi}}

By Equation (\ref{eqn1}) and (\ref{eqn2}), we expect that ``on average'', each rectangle has roughly $\delta\cdot 2^{-c_M}$ fraction 
of the pairs that are good and match the $\mml$ value of the rectangle. By Markov's inequality, we can indeed show that many rectangles 
have many \emph{columns} with at least this fraction (up to a constant factor). 
More formally, for each Merlin's message, define 
\[
	\tYpi=\left\{y\in \Ypi:\mu\left((\Xpi\times\{y\})\cap S\cap \MML^{-1}(\mml)\right)\geq\frac{\delta}{2}\cdot 2^{-c_M}\cdot \mu(\Xpi\times \{y\})\right\}
\]
to be the set of \emph{columns} in $\Rpi$ with many good input pairs and matching $\mml$,
\[
\tRpi=\Xpi\times \tYpi
\]
to be the union of these columns,
\[
\mathcal{R}_{\mml}=
\left\{\tRpi:
|\pi_A|\leq c_A, |\pi_B|\leq c_B\right\},
\]
and let
\[
\mathcal{R}=\bigcup_{\mml \; : \; |\mml|\leq c_M} \mathcal{R}_{\mml}
\]
be the set of rectangles with ``sufficiently many" good input pairs  and matching $\mml$ in \emph{every} column.

Again, by definition, for $(x, y)\in S$, when $\mml=\MML(x, y)$, the protocol outputs the correct function value $f(x, y)$. Thus, every column of each rectangle $\tRpi$ has at least $\frac{\delta}{2}\cdot 2^{-c_M}$ fraction of the inputs having the same function value, i.e., each $\tRpi$ is $\frac{\delta}{2}\cdot 2^{-c_M}$-\cm{}. 

It remains to show that at least one of these rectangles is large (satisfying propositions 2 and 3 of the lemma). 
Indeed, by Equation (\ref{eqn1}) and (\ref{eqn2}), we have
\[
	\begin{aligned}
		\sum_{\tRpi\in\mathcal{R}}\mu(\tRpi)&\geq \sum_{\tRpi\in\mathcal{R}}\mu(\tRpi\cap S\cap \MML^{-1}(\mml)) \\
		&=\sum_{\condi}\mu(\tRpi\cap S\cap \MML^{-1}(\mml)) \\
		&\geq \delta-\sum_{\condi}\mu((R_{\pi}\setminus\tRpi)\cap S\cap \MML^{-1}(\mml))  && \text{(by \eqref{eqn2})}\\
		&\geq \delta-\frac{\delta}{2}\cdot 2^{-c_M}\sum_{\condi}\mu(R_{\pi}\setminus\tRpi) && \text{(by definition of $\tRpi$)}\\
		&\geq \delta-\delta/2=\delta/2.       && \text{(by \eqref{eqn1})}
	\end{aligned}
\]
In particular, there is one $\hmml$ such that 
\begin{align} \label{eq_hmml_large}
	\sum_{\tRpi\in\mathcal{R}_{\hmml}}\mu(\tRpi)\geq (\delta/2)\cdot 2^{-c_M}.
\end{align}

From now on, let us fix Merlin's message to be $\hmml$, and focus on $\cR_{\hmml}$. 
Recall that, by definition, for each $\tRpi\in\cR_{\hmml}$ with $\pi=(\hmml,\pi_A,\pi_B)$, we have $|\pi_A|\leq c_A$ and $|\pi_B|\leq c_B$.  
For every $x\in \cX$, let 
$$ \Sx := \{\pi:\tRpi\in \cR_{\hmml},x\in \Xpi\} $$ 
be the set of all possible transcripts $\pi \in \cR_{\hmml}$ that can be generated by $x,\hmml$ (and any $y$). 
Intuitively, since Bob sends at most $c_B$ bits in $\pi$, $\Sx$ can be of size at most $2^{c_B}$. This is the content of the following 
simple claim: 
\begin{claim} \label{cl_Sx_size}
For every $x\in \cX$,  $|\Sx|\leq 2^{c_B}$. 
\end{claim}
\begin{proof}
Let $(\Pi|\cR_{\hmml})$ denote a \emph{uniformly} random transcript $\pi \in \cR_{\hmml}$. 
Since $P$ is deterministic, the random variable $(\Pi|x,\cR_{\hmml})$ is uniformly distributed over $\Sx$.
Thus, $|\Sx| = 2^{H(\Pi|x , \cR_{\hmml})}$.  
Let $\Pi_i$ denote the $i$'th message (not necessarily bit) sent in $\pi$ (assuming messages are prefix-free). 
Then we may assume, without loss of generality, that Alice speaks in odd  rounds of $\pi$ and Bob speaks in even rounds. 
By the chain rule for entropy, we have  
\begin{align*}
& H(\Pi|x , \cR_{\hmml}) = \sum_{\text{round} \; i } H(\Pi_i| \Pi_{<i} , x , \cR_{\hmml})     \\
& = \sum_{\text{even } i} H(\Pi_i| \Pi_{<i} , x , \cR_{\hmml})  \leq \sum_{\text{even } i} |(\Pi_i|\cR_{\hmml})|  \leq c_B,
\end{align*}
where in the second transition we used the fact that for messages $\Pi_i$ sent by Alice, we have $H(\Pi_i| \Pi_{<i} , x , \cR_{\hmml})=0$ 
since $\Pi$ is deterministic, and the last transition follows from the assumption that $|\pi_B|\leq c_B$ for every $\pi\in \cR_{\hmml}$. 
This completes the proof. 
\end{proof}

With this claim in hand, we can now bound the fraction of rectangles in $\cR_{\hmml}$ with a ``small Bob side" ($\tYpi$):

\begin{align} 
\sum_{\stackrel{\tRpi\in\cR_{\hmml}:}{\mu_y(\tYpi)<(\delta/4)\cdot 2^{-c_M-c_B}}}\mu(\tRpi) 
&=\sum_{\stackrel{\tRpi\in\cR_{\hmml}:}{\mu_y(\tYpi)<(\delta/4)\cdot 2^{-c_M-c_B}}}\mu_x(\Xpi)\cdot \mu_y(\tYpi)   \label{eq_lb_mu_Y_1} \\
&<\sum_{x\in\cX}\sum_{\stackrel{\tRpi\in\cR_{\hmml}:}{\mu_y(\tYpi)<(\delta/4)\cdot 2^{-c_M-c_B}}}\mu_x(x)\cdot \mathbf{1}_{\Xpi}(x)
\cdot (\delta/4)\cdot 2^{-c_M-c_B} \nonumber \\
&\leq \sum_{x\in\cX}\sum_{\pi \in \Sx}\mu_x(x)\cdot (\delta/4)\cdot 2^{-c_M-c_B} \nonumber \\
&\leq \sum_{x\in\cX}\mu_x(x)\cdot 2^{c_B}\cdot (\delta/4)\cdot 2^{-c_M-c_B} \;\;\;\;\;\;\;\;  \text{(by Claim \ref{cl_Sx_size})} \nonumber \\
& =(\delta/4)\cdot 2^{-c_M} . \label{eq_lb_mu_Y_2}
\end{align}

We now bound the fraction of rectangles in $\cR_{\hmml}$ with a ``small Alice side" ($\Xpi$). 
To this end, recall that $|\cR_{\hmml}|\leq 2^{c_A+c_B}$, and therefore

\begin{align} \label{eq_lb_mu_X}
&\sum_{\stackrel{\tRpi\in\cR_{\hmml}:}{\mu_x(\Xpi)<(\delta/4)\cdot 2^{-c_M-c_A-c_B}}}\mu(\tRpi) \nonumber \\
\leq\,\,&\sum_{\stackrel{\tRpi\in\cR_{\hmml}:}{\mu_x(\Xpi)<(\delta/4)\cdot 2^{-c_M-c_A-c_B}}}\mu_x(\Xpi) <(\delta/4)\cdot 2^{-c_M}. 
\end{align}

But by  \eqref{eq_hmml_large}, we know that  $\sum_{\tRpi\in\cR_{\hmml}}\mu(\tRpi)\geq (\delta/2)\cdot 2^{-c_M}$, 
hence there exists some $\tRpi\in\cR_{\hmml}$ such that both $\mu_y(\tYpi)\geq(\delta/4)\cdot 2^{-c_M-c_B}$ and 
$\mu_x(\Xpi)\geq(\delta/4)\cdot 2^{-c_M-c_A-c_B}$. 
\end{proof}

\begin{lemma}\label{lem_cm_rect}
Let $P$ be a $\ANC$ protocol that $\delta$-solves $f: \cX\times\cY\to \cZ$ under a product distribution 
$\mu=\mu_x\times\mu_y$.  Let $\cG_X \subseteq \cX$ and $\cG_Y \subseteq \cY$ be subsets of inputs such that 
$\Pr_{\mu}[X\in \cG_X \; \wedge \; Y\in \cG_Y] \geq 1- \eps$. Then $M(f)$ has a rectangle $R_{\pi}=X_{\pi}\times Y_{\pi}$ such that
	\begin{enumerate}
		\item
		         $\Xpi \subseteq \cG_X$ and  $\Ypi \subseteq \cG_Y$; 
		\item
			$R_{\pi}$ is $((\delta-\eps)/2)\cdot 2^{-c_M}$-column-monochromatic; 
		\item
			$\mu_x(X_{\pi})\geq (1-\eps)((\delta-\eps)/4)\cdot 2^{-(c_M+c_A+c_B)}$;
		\item
			$\mu_y(Y_{\pi})\geq (1-\eps)((\delta-\eps)/4)\cdot 2^{-(c_M+c_B)}$.
	\end{enumerate}
\end{lemma}

\begin{proof}
The claim follows directly from Lemma \ref{lem_monochromatic_rect} by considering the distribution 
$\mu' := (\mu_x|\cG_X) \times (\mu_y|\cG_Y)$. Note that $\mu'$ is still a \emph{product} distribution, 
and that $P$ must succeed in solving $f$ under $\mu'$ with probability at least $(\delta-\eps)$ (or else it will 
have success $< \delta$ under $\mu$), so we may indeed apply Lemma \ref{lem_monochromatic_rect} with $\mu'$ and 
$\delta' := \delta-\eps$ to obtain a rectangle $R = X\times Y \subseteq \cG_X\times\cG_Y$ with 
$\mu'_x(X)\geq ((\delta-\eps)/4)\cdot 2^{-(c_M+c_A+c_B)}$ and 
$\mu'_y(Y)\geq ((\delta-\eps)/4)\cdot 2^{-(c_M+c_B)}$. Finally, since 
$\mu'_x(X) \leq \mu_x(X)/\mu_x(\cG_X) \leq \mu_x(X)/(1-\epsilon)$,  it follows that 
$\mu_x(X)\geq (1-\eps) \mu_x'(X) \geq (1-\eps)((\delta-\eps)/4)\cdot 2^{-(c_M+c_A+c_B)}$. The same argument applied 
to $\mu'_y(Y)$ completes the proof.
\end{proof}

\begin{remark}[General (non-product) distributions]
The only step in the proof of Lemma \ref{lem_monochromatic_rect} that uses the independence of 
$x$ and $y$ (i.e., the product assumption on $\mu$), is  the transition in equation \eqref{eq_lb_mu_Y_1}. 
It is not hard to see that, following a similar calculation to that of Equation \eqref{eq_lb_mu_X}, it is possible to obtain a similar 
(yet weaker) lower bound on the measure of an induced rectangle  $R_{\pi}=X_{\pi}\times Y_{\pi}$ under \emph{arbitrary} (general) distributions $\mu$, namely, that 
$\mu(R) \gtrsim \delta\cdot 2^{-c_M-c_A-c_B}$. Note that such bound does not distinguish between the measure of ``Alice's side" ($X_{\pi}$) 
and ``Bob's side" ($Y_{\pi}$), so it may be less useful to ``lopsided" communication problems that typically arise from data structure reductions. 
Nevertheless, we stress that the lemma above is  more general than stated. 
\end{remark}

}

\ifx\mainfile\undefined
\lemMonochromaticRect
\lemMonochromaticRectFull

\end{document}
\else
\iftoggle{focs}{
\lemMonochromaticRectCompact
}{
\lemMonochromaticRect
\lemMonochromaticRectFull
}
\fi

\section{The Amortized Dynamic Cell-Probe Complexity of $\ORC$}\label{sec_ds_lb}

In this section, we prove our main theorem, an amortized lower bound for 2-dimensional weighted orthogonal range counting ($\ORC$) problem. 
\iftoggle{focs}{}{

\begin{restate}[Theorem~\ref{thm2d}]
\thmtwodStatement
\end{restate}
\begin{remark}
	In particular, the theorem implies the following: If $D$ probes $o(n\left(\log n/c\log\log n\right)^2)$ cells in expectation on any sequence of $O(n/\log^c n)$ updates and $O(n)$ 
	queries, then there is some operation sequence such that the probability $D$ is correct on all queries is at most $2^{-n/\log^{c+O(1)} n}$.
\end{remark}

\paragraph{Plan.} }To prove the theorem, we first define a hard distribution $\cD$ on the operation sequence for $\ORC$, and fix a data 
structure $D$. By Yao's Minimax Principle~\cite{Yao77}, we can always fix the random bits used by $D$, so that the probability that $D$ is correct on all queries and makes too few probes is preserved. We may assume $D$ is deterministic from now on.  Consider the execution of $D$ on a random sequence of operations. We shall decompose this sequence into many communication games in the $\ANC$ model, in a way that guarantees that if $D$ is fast and has decent success probability, then most of the games can be solved 
with low communication cost and non-trivial success probability. On the other hand, we prove that non of these induced games can be solved both efficiently and with non-trivial accuracy. Combining these two facts together, we conclude that 
no data structure can be fast and have decent success probability simultaneously. 

\iftoggle{focs}{}{
In the following, we first define the hard distribution $\cD$, and its corresponding communication game $G_{\ORC}$. In Section~\ref{subsect_sim_ORC}, we propose a protocol for $G_{\ORC}$ given data structure $D$. In Section~\ref{subsect_comm_lower}, we prove a lower bound for $G_{\ORC}$. In Section~\ref{subsec_dslower}, we combine the results and prove Theorem~\ref{thm2d}. 
}

\paragraph{Hard distribution $\mathcal{D}$.} The sequence always has $n/\log^c n$ updates and $n-n/\log^c n$ queries such that there are (about) $\log^c n$ queries between two consecutive updates. Every update inserts a point at a uniformly random location in the  $[n]\times[n]$ grid 
with a random weight uniformly chosen from $[n]$. Each query is a uniformly random point in the $[n]\times[n]$ grid. 
The random sequence is independent across the updates and the queries. 

More formally, let $\cD_U$ be the uniform distribution over all possible $n^3$ updates, $\cD_Q$ be the uniform distribution over all possible $n^2$ queries. Let $\cD_i$ be the distribution for $i$-th operation, i.e., $\cD_i=\cD_U$ if $i$ is multiple of $\log^c n$, and $\cD_i=\cD_Q$ otherwise. Let $\cD=\cD_1\times \cD_2\times \cdots\times \cD_n$ be our hard distribution over sequences of $n$ operations. We will focus on $\cD$ in the following.

\paragraph{The Distributional Communication Game $G_{\ORC}(k,q,n)$.} Let $\mathcal{X}$ be the set of $k$-tuples of \emph{weighted} points in $[n]\times [n]$ with weights from $[n]$, $\mathcal{Y}$ be the set of $q$-tuples of \emph{unweighted} points in $[n]\times [n]$. Let the input distribution $\mu=\mu_x\times \mu_y$ be the uniform distribution over $\mathcal{X}\times \mathcal{Y}$. Then, $x=((x_1,w_1),\ldots,(x_k,w_k))$ is a $k$-tuple of weighted points and $y=(y_1,\ldots,y_q)$ is a $q$-tuple of unweighted points. Let $\ORC(x, y): ([n]^2\times [n])^k \times ([n]^2)^q \to [kn]^q$ denote the function whose output is a $q$-tuple of numbers from $[kn]$, whose $i$-th coordinate is  the sum of $w_j$'s for which $x_j\leq y_i$,\footnote{$x_j\leq y_i$ means both coordinates of $x_j$ are no larger than the corresponding coordinates of $y_i$.} i.e., 
\[
\ORC(x, y)_i:=\sum_{j:x_j\leq y_i}w_j.
\]

\def\subsectSimORC{

Consider the communication game $G_{\ORC}(k, q, n)$ in the $\ANC$ model. Let $I_A$ and $I_B$ be two consecutive intervals in a random operation 
sequence sampled from $\cD$, such that the number of updates in $I_A$ equals to $k$ and the number of queries in $I_B$ equals to $q$, i.e., $k\sim |I_A|\cdot \log^{-c} {n}$ and $q\sim |I_B|$.\footnote{Throughout the paper, $f\sim g$ stands 
for $f=g+o(g)$ when $n$ goes to infinity.} 
We shall embed the game into a sequence of operations in the dynamic cell-probe model, so that the answers to all queries in the sequence 
produces a solution to communication game. In particular, if there is an efficient data structure $D$ for the corresponding operation sequence, 
Alice and Bob can simulate $D$
using the protocol $\SIMD$ from Section~\ref{subsect_sim_general}, which in turn would yield a too-good-to-be-true $\ANC$ protocol 
for $G_{\ORC}$. 

To this end, fix a deterministic data structure $D$, and all operations before $I_A$. Both $D$ and these operations are publicly 
known to the players and therefore can be hard-wired to the protocol. Let $P(I_A)$ and $P(I_B)$ be the set of cells probed by $D$ during $I_A$ 
and $I_B$ respectively. We now provide a $\ANC$ protocol that simulates $D$ and solves 
$G_{\ORC}(k,q,n)$.

\paragraph{The simulation protocol $P_D$ for $G_{\ORC}(k,q,n)$:}
\begin{enumerate}
	\item (Generate the operations)
		The number of updates in $I_A$ equals to $k$. The number of queries in $I_B$ equals to $q$. Alice sets $i$-th update in $I_A$ in chronological order to be \texttt{update(}$x_i$\texttt{, }$w_i$\texttt{)}. Bob sets $j$-th query in $I_B$ to be \texttt{query(}$y_j$\texttt{)}. They use public randomness to sample queries in $I_A$ and updates in $I_B$ uniformly and independently. 
	\item (Simulate $D$ on $\cO$) Let $\cO$ be the sequence of operations obtained by concatenating the hard-wired operations before $I_A$, and the operations in $I_A$ and in $I_B$ generated in the first step. Run $\SIMD$ from Section~\ref{subsect_sim_general} on $\cO$.
	\item (Bob recovers the answer to $G_{\ORC}(k,q,n)$)
		For each query in $I_B$, the answer to the query from the simulation is the sum of weights of points updated and dominated by the query. This includes the points updated before $I_A$, in $I_A$ and in $I_B$ but before the query. Bob knows exactly the updates before $I_A$ and in $I_B$. By subtracting the sum of weights of points in those time periods and dominated by the query, Bob gets the sum of weights of points updated in $I_A$ and dominated by the query. This sum is precisely the answer to the communication game.
\end{enumerate}

By Lemma~\ref{lem_sim_general}, we have the following conclusion on protocol $P_D$. 
\begin{lemma}\label{lem_sim_orc}
For protocol $P_D$, we have that
\begin{enumerate}
\item it is a valid protocol in $\ANC$;
\item Alice sends at most $4|P(I_A)\cap P(I_B)|\cdot w$ bits in expectation;
\item Bob sends at most $|P(I_A)\cap P(I_B)|\cdot \log\frac{e|P(I_A)|}{|P(I_A)\cap P(I_B)|}$ bits;
\item Merlin sends at most $|P(I_A)\cap P(I_B)|\cdot \log\frac{e|P(I_B)|}{|P(I_A)\cap P(I_B)|}+O\left(\log n\right)$ bits;
\item when the input pair $(x, y)$ are sampled from $\mu$, operations generated in $I_A$ and $I_B$ will follow distribution $\cD$;
\item the protocol outputs a correct answer for $G_{\ORC}$ if and only if $D$ is correct on all queries in $I_B$ in $\cO$. 
\end{enumerate}
\end{lemma}
}

\iftoggle{focs}{
The input of the communication game $G_{\ORC}(k,q,n)$ can be embedded into a sequence of operations for the $\ORC$ data structure problem, such that Alice's input corresponds to $k$ updates, Bob's input corresponds to $q$ queries, and output for the communication game corresponds to the answers to the $q$ queries. Assuming there is a good data structure $D$, we can solve $G_{\ORC}(k,q,n)$ efficiently using the simulation protocol in Section~\ref{subsect_sim_general}. See Appendix~\ref{app_sim_ORC} for more details.
}
{
\subsection{Efficient data-structure simulation in the $\ANC$ model}\label{subsect_sim_ORC}
\subsectSimORC
}%
\def\contlemTwoDORCcclb{
For $n$ large enough, $k\geq \sqrt{n}$ and $k/q\sim\log^{1000}n$,
$$\suc{G_\ORC}{\mu}{0.5\sqrt{kq}, 0.005q\log n, 0.0005q\log n} \leq 2^{-0.2q\log\log n}.$$ 
}%
\def\subsectCommLower{
The obvious strategy for proving this lemma is to use the argument in Section~\ref{subsect_cm_rect}, which asserts that an efficient $\ANC$ protocol implies a large \cm{} 
rectangle. Therefore, ruling out the existence of a large \cm{} rectangle in $M(G_{\ORC})$ would give us a communication lower bound on $G_{\ORC}$.  

Unfortunately, $G_{\ORC}(k,q,n)$ does in fact contain large \cm{} rectangles. For example, when all of Bob's $q$ points have $r$-coordinate smaller than $n/\log^{\Theta(1)} n$, Alice does not have to tell Bob any information about her points with $r$-coordinate greater than that quantity. Thus, in expectation, Alice only needs to speak $k/\log^{\Theta(1)} n$ bits, and this case happens with $2^{-\Theta(q\log\log n)}$ probability over a random Bob's input. In the other word, there is a \cm{} rectangle of size $2^{-k/\log^{\Theta(1)} n}\times 2^{-\Theta(q\log\log n)}$. We cannot hope to prove a communication lower bound higher than $(k/\log^{\Theta(1)} n, \Theta(q\log\log n), 0)$ using this approach alone.

\ifx\mainfile\undefined
\documentclass[11pt]{article}
\usepackage{fullpage}
\usepackage{times}
\usepackage{amsmath,amsfonts,amssymb,amsthm}
\usepackage{graphicx}
\usepackage{tikz}
\usetikzlibrary{decorations, patterns}
\newcommand{\E}{\ensuremath{\mathop{\mathbb{E}}}}
\newcommand{\cH}{\mathcal{H}}
\newcommand{\cX}{\mathcal{X}}

\newcommand{\ORC}{\mathsf{2D\text{-}ORC}}
\newcommand{\eps}{\epsilon}
\newcommand{\cm}{column-monochromatic}
\newcommand{\typ}{evenly-spreading}
\newcommand{\iso}{isolated}
\newcommand{\ws}{well-separated}
\newcommand{\tq}{\ensuremath{\tilde{Q}}}

\newcommand{\mnote}[1]{ \marginpar{\tiny\bf
            \begin{minipage}[t]{0.5in}
              \raggedright #1
           \end{minipage}}}

\begin{document}

\newtheorem{definition}{Definition}
\newtheorem{theorem}{Theorem}
\newtheorem{lemma}{Lemma}
\newtheorem{proposition}{Proposition}
\newtheorem{conjecture}{Conjecture}
\newtheorem{claim}{Claim}
\newtheorem*{remark}{Remark}

\else
\fi

\newcommand{\bgrid}{\tikz[baseline=1pt]{\fill (0,0) rectangle ++(15pt, 10pt);}}
\newcommand{\ggrid}{\tikz[baseline=1pt]{\fill [draw, fill=black!20] (0,0) rectangle ++(15pt, 10pt);}}
\newcommand{\sgrid}{\tikz[baseline=1pt]{\fill [draw, decorate, thick, pattern=north east lines] (0,0) rectangle ++(15pt, 10pt);}}
\newcommand{\cgrid}{\tikz[baseline=1pt]{\fill [draw, decorate, thick, pattern=north east lines] (0,0) rectangle ++(15pt, 10pt);\fill [decorate, thick, pattern=north west lines] (0,0) rectangle ++(15pt, 10pt);}}

To circumvent such inputs from breaking the argument, and for other technical reasons, we only consider Alice's \emph{\typ} inputs and Bob's \emph{\ws} inputs, which we will define in the following. Consider the following $B+1$ ways of partitioning $[n]^2$ into blocks of area $A$: for each $0\leq i\leq B$, $G_i$ partitions $[n]^2$ into blocks of size $\sim(A\alpha^i/n)\times (n/\alpha^{i})$. 

\begin{center}
\begin{tikzpicture}
\foreach \y in {0,5,...,60}
	\draw (0pt, \y pt) -- (60pt, \y pt);
\draw (0pt, 0pt) -- (0pt, 60pt);
\draw (60pt, 0pt) -- (60pt, 60pt);
\node at (30pt, -10pt) {$G_0$};

\foreach \y in {0,10,...,60}
	\draw (90pt, \y pt) -- (150pt, \y pt);
\foreach \x in {90,110,130,150}
	\draw (\x pt, 0pt) -- (\x pt, 60pt);
\node at (120pt, -10pt) {$G_1$};

\node at (180pt, 30pt) {\Large $\cdots$};

\foreach \y in {0,60}
	\draw (210pt, \y pt) -- (270pt, \y pt);
\foreach \x in {210,215,...,270}
	\draw (\x pt, 0pt) -- (\x pt, 60pt);
\node at (240pt, -10pt) {$G_B$};
\end{tikzpicture}
\end{center}

We set the parameters/notations in the following way:
\begin{itemize}
	\item
		let the coordinate of a point on $[n]\times [n]$ be $(r, c)$;
	\item
		let the ratio of number of Alice's points to Bob's points be $\beta=k/q\sim \log^{1000} n$.
	\item
		let the area $A\sim(4n^2\log n)/k$;
	\item
		let the ratio $\alpha\sim\beta/\log^3 n$, i.e., $A\alpha\sim 4n^2/q\log^2 n$;
	\item
		let $B=\log (n^2/A)/\log\alpha\sim\log k/997\log\log n$, so that each block in $G_B$ has size exactly $n\times A/n$. 
\end{itemize}
We ensure that $n^2/A$ is an integer and $B$ is an integer.

\begin{definition}[Evenly-spreading tuples]
We say that a $k$-tuple $S$ of points in $[n]\times [n]$ is \emph{\typ}, if in every $G_i$, all but $\leq\!\!\sqrt{kq}$ of the blocks have some points in it.
\end{definition}

The following lemma ensures that a uniformly random $S$ is evenly-spreading with extremely high probability. 
\begin{lemma}\label{lem_es_lb}
A uniformly random $k$-tuple $S$ of points is \typ{} with probability $\geq 1-2^{-q\log^{500}n}$.
\end{lemma}
\begin{proof}
For each $G_i$, the probability that it has $\geq\!\!\sqrt{kq}$ empty blocks is at most:
\[
	\begin{aligned}
		&{n^2/A \choose \sqrt{kq}}\cdot \left(1-\frac{\sqrt{kq}}{n^2/A}\right)^k \\
		\leq\,& \left(\frac{ek}{4\sqrt{kq}\log n}\right)^{\sqrt{kq}}\cdot e^{-4\sqrt{kq}\log n} \\
		\leq\,& 2^{\sqrt{kq}(500\log\log n-4\log n)} \\
		\leq\,& 2^{-q\log^{501} n}.
	\end{aligned}
\]
Hence a union bound implies that the probability $S$ is not evenly-spreading is at most:
\[
	B2^{-q\log^{501}n}\leq 2^{-q\log^{500}n}.
\]
\end{proof}

\begin{definition}[Far points]
Two points in $[n]\times [n]$ are \emph{far} from each other, if they are not in any axis-parallel rectangle of area $A\alpha$, i.e., the product of differences in two coordinates is at least $A\alpha$. 
\end{definition}

\begin{definition}[Isolated tuples]
A tuple $\tq$ of points in $[n]\times [n]$ is \emph{\iso}, if every pair of points in $\tq$ are far from each other. 
\end{definition}

\begin{definition}[Well-separated tuples]
A $q$-tuple $Q$ of points on $[n]\times [n]$ is \emph{\ws{}}, if it contains an \iso{} subtuple $\tq$  with $|\tq|\geq q/2$. 
\end{definition}

The following lemma ensures that a uniformly random $q$-tuple points is \ws{} with extremely high probability. 
\begin{lemma}\label{lem_ws_lb}
A uniformly random $q$-tuple $Q$ of points is \ws{} with probability $\geq 1-2^{-0.4q\log\log n}$.
\end{lemma}
\begin{proof}
The area of region that is not far from a given point $p_0$ can be bounded as follows:
\[
\begin{aligned}
	4\sum_{i=0}^n \min\{n,A\alpha/i\}&\leq 4(n+A\alpha\sum_{i=1}^n 1/i) \\
	&\leq 4(n+4n^2(1+\log n)/q\log^2 n) \\
	&\leq 17n^2/q\log n.
\end{aligned}
\]
Pick all points that do not land in the above region of any point \emph{before} it in the tuple. For each point, the probability that it is picked is at least $1-17\log^{-1} n$. 
The probability that no more than $q/2$ points are picked is at most:
\[
	2^q \left(17\log^{-1} n\right)^{q/2}\left(1-17\log^{-1} n\right)^q\leq 2^{-0.4q\log\log n}.
\]

\end{proof}

\begin{lemma}\label{lem_det}
Let $\tq$ be a tuple of \emph{\iso{}} points. Then 
\begin{itemize}
	\item the $r$-coordinates of all points, 
	\item the \emph{set} of blocks containing each point in each $G_i$ (for $0\leq i\leq B$), and 
	\item the $c$-offset of each point \emph{within} the block in $G_B$
\end{itemize}
together uniquely determine the locations of all points in $\tq$.
\end{lemma}
\begin{proof}
Given the $r$-coordinate of a point $p$, there are $\alpha$ blocks of $G_1$ that could contain it without providing any extra information. Since $\tq$ is \iso{}, and these $\alpha$ blocks together form a rectangle of area $A\alpha$, there can be no other points from $\tq$ in any of them. Therefore, given the set of blocks in $G_1$ containing $\tq$, exactly one of the $\alpha$ blocks will belong to this set, i.e., the block containing $p$. This reduces the range of $c$-coordinate that $p$ could be in by a factor of $\alpha$. Given this information, there are $\alpha$ blocks of $G_2$ that could contain $p$. For 
 the same exact reason, knowing the set of blocks in $G_2$ containing $\tq$ reduces the range by a factor of $\alpha$ further. 

Given all sets, we get to know for each point, the $r$-coordinate and the block in $G_B$ containing it. Therefore, knowing the offsets within $G_B$ uniquely determines the locations of all points.
\end{proof}

We are now ready to prove our main lemma for the $G_{\ORC}$ communication game, asserting that $M(G_{\ORC})$ does not contain 
any large rectangle with even a slightly better bias than the trivial one ($n^{-q}$): 

\begin{lemma}\label{lem_lb_cm_rect}
There is no rectangle $R=X\times Y$ in $M(G_{\ORC})$ satisfying all of the following conditions:
\begin{enumerate}
	\item
		every $x\in X$ is \typ{}, every $y\in Y$ is \ws{};
	\item
		$R$ is $n^{-0.001q}$-\cm;
	\item
		$\mu_x(X)\geq 2^{-\sqrt{kq}}$; 
	\item
		$\mu_y(Y)\geq 2^{-0.01q\log n}$.
\end{enumerate}
\end{lemma}

Let us first think about the case where $R$ is 1-\cm{} and $\mu_x$ is the uniform distribution over $k$ points with fixed locations (only the weights vary). In this case, all entries in every column $y\in Y$ have exactly the same function value in $R$. Given $y$ and its function value, it can be seen as imposing $q$ linear constraints to the $k$ weights, i.e., each query with the answer tells us that the sum of some points should equal to some number. Intuitively, if $|Y|$ is too large, it will be inevitable that the union of all queries appeared in $Y$ ``hit everywhere'' in $[n]\times[n]$. Even if different $q$-tuples $y\in Y$ may have overlaps in queries, it is still impossible to pack too many $y$'s in a small area. However, if the union of all queries in $Y$ hit too many places in $[n]\times [n]$, they will impose many independent linear constraints on the $k$ weights, and thus $|X|$ must be small. 

In general, we are going to focus on how many different regions of $[n]\times [n]$ $Y$ ``hits''. If it hits very few, we show that $|Y|$ must be small, by encoding each $y\in Y$ using very few bits. This encoding scheme is a variation of the encoding argument by Larsen~\cite{Larsen12a}. On the other hand, if $Y$ does hit many different regions, we show $|X|$ must be small. To show it, we will describe a new encoding scheme which encodes $x\in X$ using very few bits. The main idea is to use public randomness (which can be seen by both the encoder and the decoder, and is independent of $x$) to sample a few regions in $[n]\times [n]$, then with decent probability, $Y$ ``hits'' all these regions, and the function value is correct for some $y$ hitting some of the regions, due to the slight bias of the rectangle. They view the public randomness as infinite such samples, and the encoder just writes down the ID of the first sample that has the above property. Then the decoder will be able to ``learn'' $q$ linear equations on $x$ from it. The point here is that if the probability that a random sample has the above property is not too low, the ID of the first success sample will have less than $q\log n$ bits. Thus we will be able to use less bits compared to the naive encoding. At last, we show that if we do this multiple times, each time the $q$ linear equations the decoder learns will be independent from the previous ones with high probability. This ensures us that it is possible to apply the sampling many times to save even more bits, which would allow us to prove the lemma.

\begin{proof}
%
%

%

Consider a \ws{} $y$, and its lexicographically first\footnote{``Lexicographically first'' is only used for the unambiguity of the definition.} \iso{} $q/2$-subtuple $\tq(y)$ (we might use $\tq$ instead of $\tq(y)$ in the following, when there is no ambiguity).
In every $G_i$, the $q/2$ points in $\tq$ appear in $q/2$ different blocks due to its isolation.
Let $\mathcal{H}_i(y)$ denote the set of $q/2$ blocks in $G_i$ containing a point from $\tq$. Call $\mathcal{H}_i(y)$ the \emph{hitting pattern} of $y$ on $G_i$. 
Let $$\mathcal{H}(y) := \bigcup_{i=0}^B\mathcal{H}_i(y) $$ 
be the \emph{hitting pattern} of $y$. 
We shall see that this combinatorial object ($\mathcal{H}(y)$) captures the delicate structure of a set of $q$ queries.

Let us fix a rectangle $R=X\times Y$ that is $n^{-0.001q}$-\cm. We are going to show that either $|X|$ is small or $|Y|$ is small, as the 
lemma predicts.  To this end, consider the set $$T=\{\mathcal{H}(y):y\in Y\}$$ of hitting patterns for all $y\in Y$. 
A trivial upper bound on $|T|$ is 
\[
\begin{aligned}
	{n^2/A \choose q/2}^{B+1}&=2^{(B+1)\log {n^2/A \choose q/2}} \\
	&=2^{(B+1)q/2\cdot  \log \Theta(n^2/Aq)} \\
	&=2^{(1+o(1))\log (n^2/A)/\log \alpha\cdot q/2\cdot \log \beta} \\
	&=2^{(1+o(1))\frac{1000}{997}|\tq|\log (n^2/A)} \\
	&=2^{(1+o(1))\frac{500}{997}q\log k},
\end{aligned}
\]
simply because there can only be that many different hitting patterns in total. However, we are going to show that if
\[
	|T| \leq {n^2/A \choose q/2}^{0.95B},
\]
polynomially fewer than the potential number of patterns, then $|Y|$ must be small. Otherwise, if $|T|$ is large, then $|X|$ must be small. 

\paragraph{Case 1 : $|T|$ smaller than the threshold.} 
This case intuitively says that if $Y$ can only generate a small number of hitting-patterns, then $Y$ itself cannot be too 
large (since $\mathcal{H}_i(y)$ can be used to determine $y$ by Lemma~\ref{lem_det}). 
To formalize this intuition, we will show how to encode each $y\in Y$ using no more 
than $1.99q\log n$ bits, which implies $|Y|\leq 2^{1.99q\log n}$ and $\mu_y(Y)\leq 2^{-0.01q\log n}$.  
Given a $y\in Y$, which is a $q$-tuple of points, we apply the following encoding scheme:
\begin{enumerate}
\item
	Write down the $r$-coordinates of all points.
\item
	Write down one bit for each point in $y$, indicating whether it  belongs to the set $\tq$ or not. For each point not in $\tq$, also write down its $c$-coordinate.
\item
	Write down the hitting pattern $\mathcal{H}(y)$. 
\item
	For each point in $\tq$, write down its $c$-offset within the block in $G_B$. 
\end{enumerate}

\emph{Decoding:} The scheme above writes down the coordinates of all points not in $\tq$, so they can clearly be decoded correctly. 
For points in $\tq$, it writes down the $r$-coordinates, the set of blocks containing them in each $G_i$, and the $c$-offsets within the block in $G_B$. By Lemma~\ref{lem_det}, it uniquely determines the locations of all \ws{} points, and hence determines $y$. 

\emph{Analysis:}
Now let us estimate the number of bits it uses. Let us analyze the number of bits used in each step: 
\begin{enumerate}
\item
	It takes $q\log n$ bits to write down all the $r$-coordinates. 
\item
	It takes 1 bit for each point, and extra $\log n$ bits for each point not in $\tq$. This step takes $q+(q-|\tq|)\log n$ bits in total.
\item
	Since we know $y\in Y$, and $\mathcal{H}(y)\in T$, this step takes $\log |T|$ bits.
\item
	This step takes $|\tq|\log (A/n)$ bits. 
\end{enumerate}

When $|T|$ is smaller than the threshold, step 3 and 4 take
\[
\begin{aligned}
\log |T|+|\tq|\log (A/n)&=0.95(1+o(1))\frac{1000}{997}|\tq|\log (n^2/A)+|\tq|\log (A/n) \\
&\leq 0.955|\tq|\log (n^2/A)+|\tq|\log (A/n) \\
&=|\tq|\log n-0.045|\tq|\log (n^2/A)
\end{aligned}
\]
bits. Thus, the encoding takes
\[
\begin{aligned}
	&q\log n+(q+(q-|\tq|)\log n)+(|\tq|\log n-0.045|\tq|\log (n^2/A)) \\
	\leq\, &2q\log n+q-0.045|\tq|\log (n^2/A) \\
	\leq\, &2q\log n-0.01q\log n=1.99q\log n
\end{aligned}
\]
bits in total. We have $\mu_y(Y)= |Y|\cdot 2^{-2q\log n}\leq 2^{-0.01q\log n}$ as desired. 

\paragraph{Case 2: $|T|$ greater than the threshold. } 
The main idea for this case is to efficiently encode an $x$ assuming there is a shared random tape between the encoder and the decoder. They use the randomness to sample random hitting patterns, and hope there is one that happens to be the hitting pattern of some $y\in Y$, and $(x, y)$ is correct in $R$. Specifying one such hitting pattern is a way to identify a correct entry in row $x$ of the rectangle, which reveals some information about $x$ to the decoder. The nature of hitting patterns guarantees that independent samples with the above property reveal ``different'' information about $x$ with high probability. This allows us to encode $x$ using very few bits by repeating the above procedure multiple times, and obtain an upper bound on $|X|$.  To this end, first consider the following probabilistic argument:
\begin{enumerate}
	\item
		sample a hitting pattern over all ${n^2/A\choose q/2}^{B+1}$ possibilities uniformly at random, i.e., for each $0\leq i\leq B$, sample a set $S_i$ of $q/2$ blocks in $G_i$ uniformly and independently,
	\item
		independent of step 1, sample a uniformly random $x\in X$. 
\end{enumerate}

Then the probability $p$ that 
\begin{enumerate}
	\item
		$\exists y\in Y$, such that $S_i=\mathcal{H}_i(y)$ for all $i$, and
	\item
		the lexicographically first such $y$ has $\ORC(x, y)$-value matching the color of column $y$ of $R$,
\end{enumerate}
is at least
\[
\begin{aligned}
	p&\geq \frac{|T|}{{n^2/A \choose q/2}^{B+1}}\cdot n^{-0.001q} \\
	&\geq {n^2/A \choose q/2}^{-0.05B-1}\cdot 2^{-0.001q\log n} \\
	&=2^{-0.05(1+o(1))\frac{500}{997}q\log k-0.001q\log n} \\
	&\geq 2^{-0.026q\log k-0.001q\log n} \\
	&\geq 2^{-0.027q\log n},
\end{aligned}
\]
where the first transition is by the assumption that $R$ is $n^{-0.001q}$-column-monochromatic, and the second transition is by the 
assumption on $|T|$.
By Markov's inequality, there are at least $p/2$-fraction of the $x$'s in $X$ that with probability at least $p/2$ (over the randomn choice of hitting pattern $\{S_i\}$), 
both conditions hold.

Let this set of $x$'s be $\overline{X}$. 
We are going exhibit a randomized encoding scheme for each $x\in\overline{X}$ that uses very few bits in expectation (over the public 
randomness of the scheme). This will imply an upper bound on $|\overline{X}|$, which in turn will imply an upper bound on $|X|$. 
Given an $x\in\overline{X}$, which is a $k$-tuple of weighted points, we apply the following encoding scheme:

\begin{enumerate}
\item 
	Write down the locations of each point in $x$.
\item
	View the infinitely long public random string as infinite samples of $\{S_i\}_{0\leq i\leq B}$. Write down the index of the first sample such that $\{S_i\}\in T$, and the lexicographically first $y$ with $\mathcal{H}(y)=\{S_i\}$ satisfies that $\ORC(x, y)$ matches the color of column $y$ of $R$. This implicitly encodes the answers to $q$ queries in $y$ on $x$ for some $y$. 
\item
	Repeat Step 2 for $\sqrt{k/q}$ times (using fresh randomness each time). 
\item
	Find \emph{all} the possible $x\in \overline{X}$ that are consistent with all answers encoded in Step 2, sort them in lexicographical order. Write down the index of the $x$ that we are trying to encode in this sorted list. 
\end{enumerate}

\emph{Deocoding:} Assuming the above encoding scheme terminates and outputs an encoding, it is easy to see that we can recover each $x\in \overline{X}$. It is because in Step 4, we find all $x$'s that are still possible, and explicitly specify which one it is.

\emph{Analysis:} To show the encoding scheme uses very few bits in expectation, it will be useful to view each $x$ as a vector $v_x\in \mathbb{R}^k$, where $i$-th coordinate encodes the weight of $i$-th point 
in $x$. Since the locations of all points are written down explicitly in Step 1, each query point $y_j$ in $y$ can also be viewed as a 0-1 vector 
$u_{y_j}\in \mathbb{R}^k$, where $i$-th coordinate indicates whether the $i$-th point is \emph{dominated} by this point.\footnote{The vector $u_{y_j}$ depends on not only on $y_j$ but also on the locations of all points in $x$.}
Call $v_x$ the \emph{weight vector} of $x$, $u_{y_j}$ the \emph{dominance vector} of $y_i$.
In this notation, the sum of weights of dominated points in $x$ is just the inner product $\langle v_x,u_{y_j} \rangle $ of 
these two vectors. 
In this sense, every execution of Step 2  implicitly encodes $q$ linear constraints on $x$. Let $L$ denote the number of \emph{linearly independent}  constraints encoded in total (note that $L$ is a random variable). We have the following cost for each step:

\begin{enumerate}
\item
	It takes $2k\log n$ bits to write down all the locations.
\item
	Each time we run this step, it takes $\log 2/p\leq 0.027q\log n$ bits in expectation.
\item
	In total, all executions of Step 2 take $0.027\sqrt{kq}\log n$ bits in expectation.
\item
	Since there are $\leq n^{k-L}$ different $x$'s satisfying all linear constraints, this step takes $(k-L)\log n$ bits. 
\end{enumerate}

In order to conclude that the scheme uses a small number of bits, 
it remains show that $L$ will be large in expectation. 
This is the content of the following technical claim.

\begin{claim}
\label{cl_lin_constraints_lb}
$\E[L] \geq 0.06\sqrt{kq}$.
\end{claim}

Assuming Claim \ref{cl_lin_constraints_lb}, the expected total cost of the encoding scheme is at most 
\[
	2k\log n+0.027\sqrt{kq}\log n+(k-\E[L])\log n \leq 3k\log n-0.033\sqrt{kq}\log n.
\]

Thus, we have $|\overline{X}|\leq 2^{3k\log n-0.033\sqrt{kq}\log n}$, $\mu_x(\overline{X})=|\overline{X}|\cdot 2^{-3k\log n}\leq 2^{-0.033\sqrt{kq}\log n}$. Therefore, we have \[\mu_x(X)\leq 2^{-0.033\sqrt{kq}\log n+0.01q\log n}\ll 2^{\sqrt{kq}},\] which proves the lemma.

\begin{figure}
\begin{center}
\begin{tikzpicture}
\foreach \x in {0,30,60}
	\foreach \y in {10,30,...,90}	
		\fill [fill=black!20] (\x+15 pt, \y-10 pt) rectangle ++(15pt, 10pt);
\foreach \x in {0,30,60}
	\foreach \y in {10,30,...,70}
		\fill [decorate, thick, pattern=north east lines] (\x pt, \y pt) rectangle ++(15pt, 10pt);
\foreach \y in {0,10,...,90}
	\draw (0pt, \y pt) -- (90pt, \y pt);
\foreach \x in {0,15,...,90}
	\draw (\x pt, 0pt) -- (\x pt, 90pt);
\fill (45pt, 20pt) rectangle ++ (15pt, 10pt);
\fill [decorate, thick, pattern=north west lines] (30pt, 30pt) rectangle ++ (15pt, 10pt);
\end{tikzpicture}
\end{center}
\caption{$G_i$}\label{Gi}
\end{figure}

The only part left is to prove Claim~\ref{cl_lin_constraints_lb}. We show that each time we run Step 2, it creates many new constraints linearly independent from the previous ones with high probability. To this end, let us fix a $G_i$ and consider a block in it, e.g., the {\bgrid} block in Figure~\ref{Gi}. It is not hard to see that different query points in the block may have different dominance vectors (the 0-1 vector in $\mathbb{R}^k$). However, if we only focus on the coordinates corresponding to points from $x$ in the {\sgrid} blocks,\footnote{We may use ``{\sgrid} coordinates'' to indicate these coordinates in the following.} all query points in {\bgrid} block have the same values in these coordinates. In general, $G_i$ can be viewed as a meta-grid with $n^2/A\alpha^i$ rows and $\alpha^i$ columns. For each block in $G_i$, all dominance vectors of points in the block have the same values, in the coordinates corresponding to points in blocks with opposite row and column parities in the meta-grid (the {\sgrid} coordinates). Define the dominance vector of a block to be the dominance vector of some query point in it, restricted to all those coordinates. By the above argument, the dominance vector of a block is well-defined. Note that, different blocks in $G_i$ may have dominance vectors defined to take values in different codomains. For example, the dominance vector of a {\ggrid} or {\bgrid} block only has {\sgrid} coordinates, and vice versa. But all blocks with the same row and column parities in the meta-grid have the same codomain for their dominance vectors. 

Now let us focus on all ($\sim n^2/4A$) dominance vectors of {\ggrid} (and {\bgrid}) blocks. Except a few of them, all others are linearly independent. More specifically, we first remove all dominance vectors of blocks ($\sim \alpha^i/2$) in the first row of the meta-grid. Then by \typ{} of the input $x$, there are at most $\sqrt{kq}$ blocks in $G_i$ having no points from $x$. For every such {\sgrid} block, we remove the dominance vector of its \emph{immediate lower-right} block, e.g., if the {\cgrid} block has no points from $x$, we remove the dominance vector of the {\bgrid} block. All the remaining vectors are linearly independent. This is because if we sort all these vectors in the upper-lower left-right order of their blocks, every dominance vector has a non-zero value in some coordinate corresponding to points in its immediate upper-left block, while all previous vectors have zeros in these coordinates, i.e., every vector is independent from the vectors before it in the sorted list. In general, for all four possibilities of row and column parities in the meta-grid, among the $\sim n^2/4A$ dominance vectors, we can remove at most $\sim\sqrt{kq}+\alpha^i/2+n^2/2A\alpha^i$ of them, so that the remaining vectors are all linearly independent.

Using the above connection between the geometry of the points and linear independence, we will be able to show that each execution of step 2 creates many new linearly independent constraints with high probability. Intuitively, among the $q/2$ blocks in $S_i$, at least $q/8$ of them have the row and column parities, say they are all {\ggrid} blocks. Then we restrict all $<\sqrt{kq}$ dominance vectors from previous executions of Step 2 to the {\sgrid} coordinates. There are many independent dominance vectors among those of the {\ggrid} blocks. Thus, except with exponentially small probability, at least $q/16$ of dominance vectors of the sampled blocks will be independent from the previous ones. If there is a $y$ that $\cH(y)=\{S_i\}$, then $y$ must have one query point in each of the $q/8$ sampled {\ggrid} blocks. The dominance vector of a query point takes same values as that of the {\ggrid} block it is in in the {\sgrid} coordinates. At least $q/16$ of the query points in $y$ will create new independent linear constraints, as their dominance vectors are independent from the previous ones, even restricted to {\sgrid} coordinates. 

More formally, in Step 2 of the encoding scheme, imagine that instead of sampling $S_i$ directly, we first randomly generate the numbers of blocks in $S_i$ with different row and column parities in the meta-grid, then sample $S_i$ conditioned on these four numbers. There must be one row and column parity that has at least $q/8$ blocks in $S_i$. Without loss of generality, we assume that at least $q/8$ blocks sampled will be in odd rows and odd columns. From now on, let us focus on sampling these $\geq q/8$ blocks.

Imagine that we sample the $\geq q/8$ blocks in $S_i$ one by one independently. Before sampling each block, consider all dominance vectors of query points created by previous executions of Step 2, restricted to all coordinates corresponding to points in blocks with opposite parities (blocks in even rows and even columns), together with all dominance vectors of blocks just sampled. There are no more than $\sqrt{kq}$ such dominance vectors in total, i.e., among all $\geq n^2/4A-\sqrt{kq}-\alpha^i/2-n^2/2A\alpha^i$ independent dominance vectors, at least $\geq n^2/4A-2\sqrt{kq}-\alpha^i/2-n^2/2A\alpha^i$ of them are linearly independent from the previous ones. Thus, the dominance vector of the new sampled block is independent from the previous ones with probability at least $1-(2\sqrt{kq}+\alpha^i/2+n^2/2A\alpha^i)/(n^2/4A)$. Among the $q/8$ blocks in $S_i$, if there are at least $q/16$ of them whose dominance vectors are independent from the previous ones, and $\cH(y)=\{S_i\}$ for some $y$, then $y$ creates at least $q/16$ new independent linear constraints. Since there is one query point in each block in $S_i$, including the $q/16$ of them with dominance vector independent from the previous ones, these $q/16$ points create one independent constraint each. 

Thus, if $\cH(y)=\{S_i\}$ and $y$ imposes at most $q/16$ new independent constraint, then it must be the case that in every $S_i$, the dominance vectors of no more than $q/16$ of the blocks are independent from the previous ones. Since all $S_i$'s are sampled independently and when $0.001B\leq i\leq 0.999B$, $2\sqrt{kq}+\alpha^i/2+n^2/2A\alpha^i\leq 2\sqrt{kq}+k^{1-\Omega(1)}\leq 3\sqrt{kq}$. , the probability of the latter is at most
\[
	\begin{aligned}
		\left(\left(\frac{3\sqrt{kq}}{n^2/4A}\right)^{q/16}2^{q/8}\right)^{0.998B}&\leq \left(\left(\frac{200\sqrt{kq}}{k/\log n}\right)^{q/16}\right)^{0.998B} \\
		&\leq \left((200\beta^{-1/2}\log n)^{q/16}\right)^{0.998B} \\
		&\leq 2^{(0.998qB/16)(8-0.5\log\beta+\log\log n)} \\
		&\leq 2^{-\frac{(1+o(1))0.998q\log k\cdot 499\log\log n}{16\cdot 997\log\log n}} \\
		&\leq 2^{-0.031q\log k}.
	\end{aligned}
\]

The probability that a sample $\{S_i\}$ succeeds is at least $p/2\geq 2^{-0.026q\log k-0.001q\log n}\gg 2^{-0.031q\log k}$. Thus, each time we run step 2, with $1-o(1)$ probability, $y$ gives us at least $q/16$ new constraints. 
Therefore, the expected value of $L$ is at least $\sqrt{k/q}\cdot (1-o(1))q/16\geq 0.06\sqrt{kq}$, as claimed. 
\end{proof}

\ifx\mainfile\undefined

\end{document}

\else
\fi

\begin{proof}[Proof of Lemma~\ref{lem_2DORC_cc_lb}]
Assume there is a $\ANC$ protocol $2^{-0.2q\log\log n}$-solves $G_{\ORC}(k,q,n)$ with communication cost $(0.5\sqrt{kq}, 0.005q\log n, 0.0005q\log n)$. By Lemma~\ref{lem_es_lb} and Lemma~\ref{lem_ws_lb}, the probability that in a random input pair $(x, y)$ sampled from $\mu$, $x$ is not \typ{} or $y$ is not \ws{} is at most $1-2^{-0.4q\log\log n}-2^{-q\log^{500} n}$. By Lemma~\ref{lem_cm_rect}, for large enough $n$, $M(G_{\ORC})$ has a rectangle $R=X\times Y$ such that
\begin{enumerate}
\item
	every $x\in X$ is \typ{} and every $y\in Y$ is \ws{};
\item
	$R$ is $2^{-0.0006q\log n}$-\cm;
\item
	$\mu_x(X)\geq 2^{-0.6\sqrt{kq}}$;
\item
	$\mu_y(Y)\geq 2^{-0.006q\log n}$.
\end{enumerate}

However, by Lemma~\ref{lem_lb_cm_rect}, such rectangle cannot exist in $M(G_{\ORC})$. We have a contradiction. 
\end{proof}
}%
\iftoggle{focs}{On the other hand, the following lemma asserts that }{
\subsection{$\ANC$ communication complexity of orthogonal range counting}\label{subsect_comm_lower}

In this section we prove Lemma~\ref{lem_2DORC_cc_lb}, asserting that }the probability of any $\ANC$ protocol with communication 
$(o(\sqrt{kq}), o(q\log n), o(q\log n))$ in solving all $q$ queries of $G_{\ORC}$ correctly, is hardly any better than the trivial 
probability obtained by randomly guessing the answers.
\iftoggle{focs}{An intuitive overview and the formal proof of this lemma can be found in Appendix~\ref{app_comm_lower}.}

\begin{lemma}[``Direct Product'' for $\ORC$ in the $\ANC$ model]\label{lem_2DORC_cc_lb}
\contlemTwoDORCcclb
\end{lemma}

\iftoggle{focs}{}{
The obvious strategy for proving this lemma is to use the argument in Section~\ref{subsect_cm_rect}, which asserts that an efficient $\ANC$ protocol implies a large \cm{} 
rectangle. Therefore, ruling out the existence of a large \cm{} rectangle in $M(G_{\ORC})$ would give us a communication lower bound on $G_{\ORC}$.  

Unfortunately, $G_{\ORC}(k,q,n)$ does in fact contain large \cm{} rectangles. For example, when all of Bob's $q$ points have $r$-coordinate smaller than $n/\log^{\Theta(1)} n$, Alice does not have to tell Bob any information about her points with $r$-coordinate greater than that quantity. Thus, in expectation, Alice only needs to speak $k/\log^{\Theta(1)} n$ bits, and this case happens with $2^{-\Theta(q\log\log n)}$ probability over a random Bob's input. In the other word, there is a \cm{} rectangle of size $2^{-k/\log^{\Theta(1)} n}\times 2^{-\Theta(q\log\log n)}$. We cannot hope to prove a communication lower bound higher than $(k/\log^{\Theta(1)} n, \Theta(q\log\log n), 0)$ using this approach alone.

\ifx\mainfile\undefined
\documentclass[11pt]{article}
\usepackage{fullpage}
\usepackage{times}
\usepackage{amsmath,amsfonts,amssymb,amsthm}
\usepackage{graphicx}
\usepackage{tikz}
\usetikzlibrary{decorations, patterns}
\newcommand{\E}{\ensuremath{\mathop{\mathbb{E}}}}
\newcommand{\cH}{\mathcal{H}}
\newcommand{\cX}{\mathcal{X}}

\newcommand{\ORC}{\mathsf{2D\text{-}ORC}}
\newcommand{\eps}{\epsilon}
\newcommand{\cm}{column-monochromatic}
\newcommand{\typ}{evenly-spreading}
\newcommand{\iso}{isolated}
\newcommand{\ws}{well-separated}
\newcommand{\tq}{\ensuremath{\tilde{Q}}}

\newcommand{\mnote}[1]{ \marginpar{\tiny\bf
            \begin{minipage}[t]{0.5in}
              \raggedright #1
           \end{minipage}}}

\begin{document}

\newtheorem{definition}{Definition}
\newtheorem{theorem}{Theorem}
\newtheorem{lemma}{Lemma}
\newtheorem{proposition}{Proposition}
\newtheorem{conjecture}{Conjecture}
\newtheorem{claim}{Claim}
\newtheorem*{remark}{Remark}

\else
\fi

\newcommand{\bgrid}{\tikz[baseline=1pt]{\fill (0,0) rectangle ++(15pt, 10pt);}}
\newcommand{\ggrid}{\tikz[baseline=1pt]{\fill [draw, fill=black!20] (0,0) rectangle ++(15pt, 10pt);}}
\newcommand{\sgrid}{\tikz[baseline=1pt]{\fill [draw, decorate, thick, pattern=north east lines] (0,0) rectangle ++(15pt, 10pt);}}
\newcommand{\cgrid}{\tikz[baseline=1pt]{\fill [draw, decorate, thick, pattern=north east lines] (0,0) rectangle ++(15pt, 10pt);\fill [decorate, thick, pattern=north west lines] (0,0) rectangle ++(15pt, 10pt);}}

To circumvent such inputs from breaking the argument, and for other technical reasons, we only consider Alice's \emph{\typ} inputs and Bob's \emph{\ws} inputs, which we will define in the following. Consider the following $B+1$ ways of partitioning $[n]^2$ into blocks of area $A$: for each $0\leq i\leq B$, $G_i$ partitions $[n]^2$ into blocks of size $\sim(A\alpha^i/n)\times (n/\alpha^{i})$. 

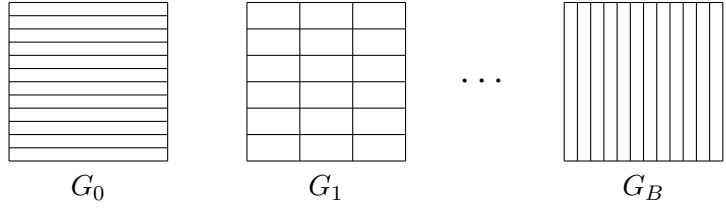
\begin{center}
\begin{tikzpicture}
\foreach \y in {0,5,...,60}
	\draw (0pt, \y pt) -- (60pt, \y pt);
\draw (0pt, 0pt) -- (0pt, 60pt);
\draw (60pt, 0pt) -- (60pt, 60pt);
\node at (30pt, -10pt) {$G_0$};

\foreach \y in {0,10,...,60}
	\draw (90pt, \y pt) -- (150pt, \y pt);
\foreach \x in {90,110,130,150}
	\draw (\x pt, 0pt) -- (\x pt, 60pt);
\node at (120pt, -10pt) {$G_1$};

\node at (180pt, 30pt) {\Large $\cdots$};

\foreach \y in {0,60}
	\draw (210pt, \y pt) -- (270pt, \y pt);
\foreach \x in {210,215,...,270}
	\draw (\x pt, 0pt) -- (\x pt, 60pt);
\node at (240pt, -10pt) {$G_B$};
\end{tikzpicture}
\end{center}

We set the parameters/notations in the following way:
\begin{itemize}
	\item
		let the coordinate of a point on $[n]\times [n]$ be $(r, c)$;
	\item
		let the ratio of number of Alice's points to Bob's points be $\beta=k/q\sim \log^{1000} n$.
	\item
		let the area $A\sim(4n^2\log n)/k$;
	\item
		let the ratio $\alpha\sim\beta/\log^3 n$, i.e., $A\alpha\sim 4n^2/q\log^2 n$;
	\item
		let $B=\log (n^2/A)/\log\alpha\sim\log k/997\log\log n$, so that each block in $G_B$ has size exactly $n\times A/n$. 
\end{itemize}
We ensure that $n^2/A$ is an integer and $B$ is an integer.

\begin{definition}[Evenly-spreading tuples]
We say that a $k$-tuple $S$ of points in $[n]\times [n]$ is \emph{\typ}, if in every $G_i$, all but $\leq\!\!\sqrt{kq}$ of the blocks have some points in it.
\end{definition}

The following lemma ensures that a uniformly random $S$ is evenly-spreading with extremely high probability. 
\begin{lemma}\label{lem_es_lb}
A uniformly random $k$-tuple $S$ of points is \typ{} with probability $\geq 1-2^{-q\log^{500}n}$.
\end{lemma}
\begin{proof}
For each $G_i$, the probability that it has $\geq\!\!\sqrt{kq}$ empty blocks is at most:
\[
	\begin{aligned}
		&{n^2/A \choose \sqrt{kq}}\cdot \left(1-\frac{\sqrt{kq}}{n^2/A}\right)^k \\
		\leq\,& \left(\frac{ek}{4\sqrt{kq}\log n}\right)^{\sqrt{kq}}\cdot e^{-4\sqrt{kq}\log n} \\
		\leq\,& 2^{\sqrt{kq}(500\log\log n-4\log n)} \\
		\leq\,& 2^{-q\log^{501} n}.
	\end{aligned}
\]
Hence a union bound implies that the probability $S$ is not evenly-spreading is at most:
\[
	B2^{-q\log^{501}n}\leq 2^{-q\log^{500}n}.
\]
\end{proof}

\begin{definition}[Far points]
Two points in $[n]\times [n]$ are \emph{far} from each other, if they are not in any axis-parallel rectangle of area $A\alpha$, i.e., the product of differences in two coordinates is at least $A\alpha$. 
\end{definition}

\begin{definition}[Isolated tuples]
A tuple $\tq$ of points in $[n]\times [n]$ is \emph{\iso}, if every pair of points in $\tq$ are far from each other. 
\end{definition}

\begin{definition}[Well-separated tuples]
A $q$-tuple $Q$ of points on $[n]\times [n]$ is \emph{\ws{}}, if it contains an \iso{} subtuple $\tq$  with $|\tq|\geq q/2$. 
\end{definition}

The following lemma ensures that a uniformly random $q$-tuple points is \ws{} with extremely high probability. 
\begin{lemma}\label{lem_ws_lb}
A uniformly random $q$-tuple $Q$ of points is \ws{} with probability $\geq 1-2^{-0.4q\log\log n}$.
\end{lemma}
\begin{proof}
The area of region that is not far from a given point $p_0$ can be bounded as follows:
\[
\begin{aligned}
	4\sum_{i=0}^n \min\{n,A\alpha/i\}&\leq 4(n+A\alpha\sum_{i=1}^n 1/i) \\
	&\leq 4(n+4n^2(1+\log n)/q\log^2 n) \\
	&\leq 17n^2/q\log n.
\end{aligned}
\]
Pick all points that do not land in the above region of any point \emph{before} it in the tuple. For each point, the probability that it is picked is at least $1-17\log^{-1} n$. 
The probability that no more than $q/2$ points are picked is at most:
\[
	2^q \left(17\log^{-1} n\right)^{q/2}\left(1-17\log^{-1} n\right)^q\leq 2^{-0.4q\log\log n}.
\]

\end{proof}

\begin{lemma}\label{lem_det}
Let $\tq$ be a tuple of \emph{\iso{}} points. Then 
\begin{itemize}
	\item the $r$-coordinates of all points, 
	\item the \emph{set} of blocks containing each point in each $G_i$ (for $0\leq i\leq B$), and 
	\item the $c$-offset of each point \emph{within} the block in $G_B$
\end{itemize}
together uniquely determine the locations of all points in $\tq$.
\end{lemma}
\begin{proof}
Given the $r$-coordinate of a point $p$, there are $\alpha$ blocks of $G_1$ that could contain it without providing any extra information. Since $\tq$ is \iso{}, and these $\alpha$ blocks together form a rectangle of area $A\alpha$, there can be no other points from $\tq$ in any of them. Therefore, given the set of blocks in $G_1$ containing $\tq$, exactly one of the $\alpha$ blocks will belong to this set, i.e., the block containing $p$. This reduces the range of $c$-coordinate that $p$ could be in by a factor of $\alpha$. Given this information, there are $\alpha$ blocks of $G_2$ that could contain $p$. For 
 the same exact reason, knowing the set of blocks in $G_2$ containing $\tq$ reduces the range by a factor of $\alpha$ further. 

Given all sets, we get to know for each point, the $r$-coordinate and the block in $G_B$ containing it. Therefore, knowing the offsets within $G_B$ uniquely determines the locations of all points.
\end{proof}

We are now ready to prove our main lemma for the $G_{\ORC}$ communication game, asserting that $M(G_{\ORC})$ does not contain 
any large rectangle with even a slightly better bias than the trivial one ($n^{-q}$): 

\begin{lemma}\label{lem_lb_cm_rect}
There is no rectangle $R=X\times Y$ in $M(G_{\ORC})$ satisfying all of the following conditions:
\begin{enumerate}
	\item
		every $x\in X$ is \typ{}, every $y\in Y$ is \ws{};
	\item
		$R$ is $n^{-0.001q}$-\cm;
	\item
		$\mu_x(X)\geq 2^{-\sqrt{kq}}$; 
	\item
		$\mu_y(Y)\geq 2^{-0.01q\log n}$.
\end{enumerate}
\end{lemma}

Let us first think about the case where $R$ is 1-\cm{} and $\mu_x$ is the uniform distribution over $k$ points with fixed locations (only the weights vary). In this case, all entries in every column $y\in Y$ have exactly the same function value in $R$. Given $y$ and its function value, it can be seen as imposing $q$ linear constraints to the $k$ weights, i.e., each query with the answer tells us that the sum of some points should equal to some number. Intuitively, if $|Y|$ is too large, it will be inevitable that the union of all queries appeared in $Y$ ``hit everywhere'' in $[n]\times[n]$. Even if different $q$-tuples $y\in Y$ may have overlaps in queries, it is still impossible to pack too many $y$'s in a small area. However, if the union of all queries in $Y$ hit too many places in $[n]\times [n]$, they will impose many independent linear constraints on the $k$ weights, and thus $|X|$ must be small. 

In general, we are going to focus on how many different regions of $[n]\times [n]$ $Y$ ``hits''. If it hits very few, we show that $|Y|$ must be small, by encoding each $y\in Y$ using very few bits. This encoding scheme is a variation of the encoding argument by Larsen~\cite{Larsen12a}. On the other hand, if $Y$ does hit many different regions, we show $|X|$ must be small. To show it, we will describe a new encoding scheme which encodes $x\in X$ using very few bits. The main idea is to use public randomness (which can be seen by both the encoder and the decoder, and is independent of $x$) to sample a few regions in $[n]\times [n]$, then with decent probability, $Y$ ``hits'' all these regions, and the function value is correct for some $y$ hitting some of the regions, due to the slight bias of the rectangle. They view the public randomness as infinite such samples, and the encoder just writes down the ID of the first sample that has the above property. Then the decoder will be able to ``learn'' $q$ linear equations on $x$ from it. The point here is that if the probability that a random sample has the above property is not too low, the ID of the first success sample will have less than $q\log n$ bits. Thus we will be able to use less bits compared to the naive encoding. At last, we show that if we do this multiple times, each time the $q$ linear equations the decoder learns will be independent from the previous ones with high probability. This ensures us that it is possible to apply the sampling many times to save even more bits, which would allow us to prove the lemma.

\begin{proof}
%
%

%

Consider a \ws{} $y$, and its lexicographically first\footnote{``Lexicographically first'' is only used for the unambiguity of the definition.} \iso{} $q/2$-subtuple $\tq(y)$ (we might use $\tq$ instead of $\tq(y)$ in the following, when there is no ambiguity).
In every $G_i$, the $q/2$ points in $\tq$ appear in $q/2$ different blocks due to its isolation.
Let $\mathcal{H}_i(y)$ denote the set of $q/2$ blocks in $G_i$ containing a point from $\tq$. Call $\mathcal{H}_i(y)$ the \emph{hitting pattern} of $y$ on $G_i$. 
Let $$\mathcal{H}(y) := \bigcup_{i=0}^B\mathcal{H}_i(y) $$ 
be the \emph{hitting pattern} of $y$. 
We shall see that this combinatorial object ($\mathcal{H}(y)$) captures the delicate structure of a set of $q$ queries.

Let us fix a rectangle $R=X\times Y$ that is $n^{-0.001q}$-\cm. We are going to show that either $|X|$ is small or $|Y|$ is small, as the 
lemma predicts.  To this end, consider the set $$T=\{\mathcal{H}(y):y\in Y\}$$ of hitting patterns for all $y\in Y$. 
A trivial upper bound on $|T|$ is 
\[
\begin{aligned}
	{n^2/A \choose q/2}^{B+1}&=2^{(B+1)\log {n^2/A \choose q/2}} \\
	&=2^{(B+1)q/2\cdot  \log \Theta(n^2/Aq)} \\
	&=2^{(1+o(1))\log (n^2/A)/\log \alpha\cdot q/2\cdot \log \beta} \\
	&=2^{(1+o(1))\frac{1000}{997}|\tq|\log (n^2/A)} \\
	&=2^{(1+o(1))\frac{500}{997}q\log k},
\end{aligned}
\]
simply because there can only be that many different hitting patterns in total. However, we are going to show that if
\[
	|T| \leq {n^2/A \choose q/2}^{0.95B},
\]
polynomially fewer than the potential number of patterns, then $|Y|$ must be small. Otherwise, if $|T|$ is large, then $|X|$ must be small. 

\paragraph{Case 1 : $|T|$ smaller than the threshold.} 
This case intuitively says that if $Y$ can only generate a small number of hitting-patterns, then $Y$ itself cannot be too 
large (since $\mathcal{H}_i(y)$ can be used to determine $y$ by Lemma~\ref{lem_det}). 
To formalize this intuition, we will show how to encode each $y\in Y$ using no more 
than $1.99q\log n$ bits, which implies $|Y|\leq 2^{1.99q\log n}$ and $\mu_y(Y)\leq 2^{-0.01q\log n}$.  
Given a $y\in Y$, which is a $q$-tuple of points, we apply the following encoding scheme:
\begin{enumerate}
\item
	Write down the $r$-coordinates of all points.
\item
	Write down one bit for each point in $y$, indicating whether it  belongs to the set $\tq$ or not. For each point not in $\tq$, also write down its $c$-coordinate.
\item
	Write down the hitting pattern $\mathcal{H}(y)$. 
\item
	For each point in $\tq$, write down its $c$-offset within the block in $G_B$. 
\end{enumerate}

\emph{Decoding:} The scheme above writes down the coordinates of all points not in $\tq$, so they can clearly be decoded correctly. 
For points in $\tq$, it writes down the $r$-coordinates, the set of blocks containing them in each $G_i$, and the $c$-offsets within the block in $G_B$. By Lemma~\ref{lem_det}, it uniquely determines the locations of all \ws{} points, and hence determines $y$. 

\emph{Analysis:}
Now let us estimate the number of bits it uses. Let us analyze the number of bits used in each step: 
\begin{enumerate}
\item
	It takes $q\log n$ bits to write down all the $r$-coordinates. 
\item
	It takes 1 bit for each point, and extra $\log n$ bits for each point not in $\tq$. This step takes $q+(q-|\tq|)\log n$ bits in total.
\item
	Since we know $y\in Y$, and $\mathcal{H}(y)\in T$, this step takes $\log |T|$ bits.
\item
	This step takes $|\tq|\log (A/n)$ bits. 
\end{enumerate}

When $|T|$ is smaller than the threshold, step 3 and 4 take
\[
\begin{aligned}
\log |T|+|\tq|\log (A/n)&=0.95(1+o(1))\frac{1000}{997}|\tq|\log (n^2/A)+|\tq|\log (A/n) \\
&\leq 0.955|\tq|\log (n^2/A)+|\tq|\log (A/n) \\
&=|\tq|\log n-0.045|\tq|\log (n^2/A)
\end{aligned}
\]
bits. Thus, the encoding takes
\[
\begin{aligned}
	&q\log n+(q+(q-|\tq|)\log n)+(|\tq|\log n-0.045|\tq|\log (n^2/A)) \\
	\leq\, &2q\log n+q-0.045|\tq|\log (n^2/A) \\
	\leq\, &2q\log n-0.01q\log n=1.99q\log n
\end{aligned}
\]
bits in total. We have $\mu_y(Y)= |Y|\cdot 2^{-2q\log n}\leq 2^{-0.01q\log n}$ as desired. 

\paragraph{Case 2: $|T|$ greater than the threshold. } 
The main idea for this case is to efficiently encode an $x$ assuming there is a shared random tape between the encoder and the decoder. They use the randomness to sample random hitting patterns, and hope there is one that happens to be the hitting pattern of some $y\in Y$, and $(x, y)$ is correct in $R$. Specifying one such hitting pattern is a way to identify a correct entry in row $x$ of the rectangle, which reveals some information about $x$ to the decoder. The nature of hitting patterns guarantees that independent samples with the above property reveal ``different'' information about $x$ with high probability. This allows us to encode $x$ using very few bits by repeating the above procedure multiple times, and obtain an upper bound on $|X|$.  To this end, first consider the following probabilistic argument:
\begin{enumerate}
	\item
		sample a hitting pattern over all ${n^2/A\choose q/2}^{B+1}$ possibilities uniformly at random, i.e., for each $0\leq i\leq B$, sample a set $S_i$ of $q/2$ blocks in $G_i$ uniformly and independently,
	\item
		independent of step 1, sample a uniformly random $x\in X$. 
\end{enumerate}

Then the probability $p$ that 
\begin{enumerate}
	\item
		$\exists y\in Y$, such that $S_i=\mathcal{H}_i(y)$ for all $i$, and
	\item
		the lexicographically first such $y$ has $\ORC(x, y)$-value matching the color of column $y$ of $R$,
\end{enumerate}
is at least
\[
\begin{aligned}
	p&\geq \frac{|T|}{{n^2/A \choose q/2}^{B+1}}\cdot n^{-0.001q} \\
	&\geq {n^2/A \choose q/2}^{-0.05B-1}\cdot 2^{-0.001q\log n} \\
	&=2^{-0.05(1+o(1))\frac{500}{997}q\log k-0.001q\log n} \\
	&\geq 2^{-0.026q\log k-0.001q\log n} \\
	&\geq 2^{-0.027q\log n},
\end{aligned}
\]
where the first transition is by the assumption that $R$ is $n^{-0.001q}$-column-monochromatic, and the second transition is by the 
assumption on $|T|$.
By Markov's inequality, there are at least $p/2$-fraction of the $x$'s in $X$ that with probability at least $p/2$ (over the randomn choice of hitting pattern $\{S_i\}$), 
both conditions hold.

Let this set of $x$'s be $\overline{X}$. 
We are going exhibit a randomized encoding scheme for each $x\in\overline{X}$ that uses very few bits in expectation (over the public 
randomness of the scheme). This will imply an upper bound on $|\overline{X}|$, which in turn will imply an upper bound on $|X|$. 
Given an $x\in\overline{X}$, which is a $k$-tuple of weighted points, we apply the following encoding scheme:

\begin{enumerate}
\item 
	Write down the locations of each point in $x$.
\item
	View the infinitely long public random string as infinite samples of $\{S_i\}_{0\leq i\leq B}$. Write down the index of the first sample such that $\{S_i\}\in T$, and the lexicographically first $y$ with $\mathcal{H}(y)=\{S_i\}$ satisfies that $\ORC(x, y)$ matches the color of column $y$ of $R$. This implicitly encodes the answers to $q$ queries in $y$ on $x$ for some $y$. 
\item
	Repeat Step 2 for $\sqrt{k/q}$ times (using fresh randomness each time). 
\item
	Find \emph{all} the possible $x\in \overline{X}$ that are consistent with all answers encoded in Step 2, sort them in lexicographical order. Write down the index of the $x$ that we are trying to encode in this sorted list. 
\end{enumerate}

\emph{Deocoding:} Assuming the above encoding scheme terminates and outputs an encoding, it is easy to see that we can recover each $x\in \overline{X}$. It is because in Step 4, we find all $x$'s that are still possible, and explicitly specify which one it is.

\emph{Analysis:} To show the encoding scheme uses very few bits in expectation, it will be useful to view each $x$ as a vector $v_x\in \mathbb{R}^k$, where $i$-th coordinate encodes the weight of $i$-th point 
in $x$. Since the locations of all points are written down explicitly in Step 1, each query point $y_j$ in $y$ can also be viewed as a 0-1 vector 
$u_{y_j}\in \mathbb{R}^k$, where $i$-th coordinate indicates whether the $i$-th point is \emph{dominated} by this point.\footnote{The vector $u_{y_j}$ depends on not only on $y_j$ but also on the locations of all points in $x$.}
Call $v_x$ the \emph{weight vector} of $x$, $u_{y_j}$ the \emph{dominance vector} of $y_i$.
In this notation, the sum of weights of dominated points in $x$ is just the inner product $\langle v_x,u_{y_j} \rangle $ of 
these two vectors. 
In this sense, every execution of Step 2  implicitly encodes $q$ linear constraints on $x$. Let $L$ denote the number of \emph{linearly independent}  constraints encoded in total (note that $L$ is a random variable). We have the following cost for each step:

\begin{enumerate}
\item
	It takes $2k\log n$ bits to write down all the locations.
\item
	Each time we run this step, it takes $\log 2/p\leq 0.027q\log n$ bits in expectation.
\item
	In total, all executions of Step 2 take $0.027\sqrt{kq}\log n$ bits in expectation.
\item
	Since there are $\leq n^{k-L}$ different $x$'s satisfying all linear constraints, this step takes $(k-L)\log n$ bits. 
\end{enumerate}

In order to conclude that the scheme uses a small number of bits, 
it remains show that $L$ will be large in expectation. 
This is the content of the following technical claim.

\begin{claim}
\label{cl_lin_constraints_lb}
$\E[L] \geq 0.06\sqrt{kq}$.
\end{claim}

Assuming Claim \ref{cl_lin_constraints_lb}, the expected total cost of the encoding scheme is at most 
\[
	2k\log n+0.027\sqrt{kq}\log n+(k-\E[L])\log n \leq 3k\log n-0.033\sqrt{kq}\log n.
\]

Thus, we have $|\overline{X}|\leq 2^{3k\log n-0.033\sqrt{kq}\log n}$, $\mu_x(\overline{X})=|\overline{X}|\cdot 2^{-3k\log n}\leq 2^{-0.033\sqrt{kq}\log n}$. Therefore, we have \[\mu_x(X)\leq 2^{-0.033\sqrt{kq}\log n+0.01q\log n}\ll 2^{\sqrt{kq}},\] which proves the lemma.

\begin{figure}
\begin{center}
\begin{tikzpicture}
\foreach \x in {0,30,60}
	\foreach \y in {10,30,...,90}	
		\fill [fill=black!20] (\x+15 pt, \y-10 pt) rectangle ++(15pt, 10pt);
\foreach \x in {0,30,60}
	\foreach \y in {10,30,...,70}
		\fill [decorate, thick, pattern=north east lines] (\x pt, \y pt) rectangle ++(15pt, 10pt);
\foreach \y in {0,10,...,90}
	\draw (0pt, \y pt) -- (90pt, \y pt);
\foreach \x in {0,15,...,90}
	\draw (\x pt, 0pt) -- (\x pt, 90pt);
\fill (45pt, 20pt) rectangle ++ (15pt, 10pt);
\fill [decorate, thick, pattern=north west lines] (30pt, 30pt) rectangle ++ (15pt, 10pt);
\end{tikzpicture}
\end{center}
\caption{$G_i$}\label{Gi}
\end{figure}

The only part left is to prove Claim~\ref{cl_lin_constraints_lb}. We show that each time we run Step 2, it creates many new constraints linearly independent from the previous ones with high probability. To this end, let us fix a $G_i$ and consider a block in it, e.g., the {\bgrid} block in Figure~\ref{Gi}. It is not hard to see that different query points in the block may have different dominance vectors (the 0-1 vector in $\mathbb{R}^k$). However, if we only focus on the coordinates corresponding to points from $x$ in the {\sgrid} blocks,\footnote{We may use ``{\sgrid} coordinates'' to indicate these coordinates in the following.} all query points in {\bgrid} block have the same values in these coordinates. In general, $G_i$ can be viewed as a meta-grid with $n^2/A\alpha^i$ rows and $\alpha^i$ columns. For each block in $G_i$, all dominance vectors of points in the block have the same values, in the coordinates corresponding to points in blocks with opposite row and column parities in the meta-grid (the {\sgrid} coordinates). Define the dominance vector of a block to be the dominance vector of some query point in it, restricted to all those coordinates. By the above argument, the dominance vector of a block is well-defined. Note that, different blocks in $G_i$ may have dominance vectors defined to take values in different codomains. For example, the dominance vector of a {\ggrid} or {\bgrid} block only has {\sgrid} coordinates, and vice versa. But all blocks with the same row and column parities in the meta-grid have the same codomain for their dominance vectors. 

Now let us focus on all ($\sim n^2/4A$) dominance vectors of {\ggrid} (and {\bgrid}) blocks. Except a few of them, all others are linearly independent. More specifically, we first remove all dominance vectors of blocks ($\sim \alpha^i/2$) in the first row of the meta-grid. Then by \typ{} of the input $x$, there are at most $\sqrt{kq}$ blocks in $G_i$ having no points from $x$. For every such {\sgrid} block, we remove the dominance vector of its \emph{immediate lower-right} block, e.g., if the {\cgrid} block has no points from $x$, we remove the dominance vector of the {\bgrid} block. All the remaining vectors are linearly independent. This is because if we sort all these vectors in the upper-lower left-right order of their blocks, every dominance vector has a non-zero value in some coordinate corresponding to points in its immediate upper-left block, while all previous vectors have zeros in these coordinates, i.e., every vector is independent from the vectors before it in the sorted list. In general, for all four possibilities of row and column parities in the meta-grid, among the $\sim n^2/4A$ dominance vectors, we can remove at most $\sim\sqrt{kq}+\alpha^i/2+n^2/2A\alpha^i$ of them, so that the remaining vectors are all linearly independent.

Using the above connection between the geometry of the points and linear independence, we will be able to show that each execution of step 2 creates many new linearly independent constraints with high probability. Intuitively, among the $q/2$ blocks in $S_i$, at least $q/8$ of them have the row and column parities, say they are all {\ggrid} blocks. Then we restrict all $<\sqrt{kq}$ dominance vectors from previous executions of Step 2 to the {\sgrid} coordinates. There are many independent dominance vectors among those of the {\ggrid} blocks. Thus, except with exponentially small probability, at least $q/16$ of dominance vectors of the sampled blocks will be independent from the previous ones. If there is a $y$ that $\cH(y)=\{S_i\}$, then $y$ must have one query point in each of the $q/8$ sampled {\ggrid} blocks. The dominance vector of a query point takes same values as that of the {\ggrid} block it is in in the {\sgrid} coordinates. At least $q/16$ of the query points in $y$ will create new independent linear constraints, as their dominance vectors are independent from the previous ones, even restricted to {\sgrid} coordinates. 

More formally, in Step 2 of the encoding scheme, imagine that instead of sampling $S_i$ directly, we first randomly generate the numbers of blocks in $S_i$ with different row and column parities in the meta-grid, then sample $S_i$ conditioned on these four numbers. There must be one row and column parity that has at least $q/8$ blocks in $S_i$. Without loss of generality, we assume that at least $q/8$ blocks sampled will be in odd rows and odd columns. From now on, let us focus on sampling these $\geq q/8$ blocks.

Imagine that we sample the $\geq q/8$ blocks in $S_i$ one by one independently. Before sampling each block, consider all dominance vectors of query points created by previous executions of Step 2, restricted to all coordinates corresponding to points in blocks with opposite parities (blocks in even rows and even columns), together with all dominance vectors of blocks just sampled. There are no more than $\sqrt{kq}$ such dominance vectors in total, i.e., among all $\geq n^2/4A-\sqrt{kq}-\alpha^i/2-n^2/2A\alpha^i$ independent dominance vectors, at least $\geq n^2/4A-2\sqrt{kq}-\alpha^i/2-n^2/2A\alpha^i$ of them are linearly independent from the previous ones. Thus, the dominance vector of the new sampled block is independent from the previous ones with probability at least $1-(2\sqrt{kq}+\alpha^i/2+n^2/2A\alpha^i)/(n^2/4A)$. Among the $q/8$ blocks in $S_i$, if there are at least $q/16$ of them whose dominance vectors are independent from the previous ones, and $\cH(y)=\{S_i\}$ for some $y$, then $y$ creates at least $q/16$ new independent linear constraints. Since there is one query point in each block in $S_i$, including the $q/16$ of them with dominance vector independent from the previous ones, these $q/16$ points create one independent constraint each. 

Thus, if $\cH(y)=\{S_i\}$ and $y$ imposes at most $q/16$ new independent constraint, then it must be the case that in every $S_i$, the dominance vectors of no more than $q/16$ of the blocks are independent from the previous ones. Since all $S_i$'s are sampled independently and when $0.001B\leq i\leq 0.999B$, $2\sqrt{kq}+\alpha^i/2+n^2/2A\alpha^i\leq 2\sqrt{kq}+k^{1-\Omega(1)}\leq 3\sqrt{kq}$. , the probability of the latter is at most
\[
	\begin{aligned}
		\left(\left(\frac{3\sqrt{kq}}{n^2/4A}\right)^{q/16}2^{q/8}\right)^{0.998B}&\leq \left(\left(\frac{200\sqrt{kq}}{k/\log n}\right)^{q/16}\right)^{0.998B} \\
		&\leq \left((200\beta^{-1/2}\log n)^{q/16}\right)^{0.998B} \\
		&\leq 2^{(0.998qB/16)(8-0.5\log\beta+\log\log n)} \\
		&\leq 2^{-\frac{(1+o(1))0.998q\log k\cdot 499\log\log n}{16\cdot 997\log\log n}} \\
		&\leq 2^{-0.031q\log k}.
	\end{aligned}
\]

The probability that a sample $\{S_i\}$ succeeds is at least $p/2\geq 2^{-0.026q\log k-0.001q\log n}\gg 2^{-0.031q\log k}$. Thus, each time we run step 2, with $1-o(1)$ probability, $y$ gives us at least $q/16$ new constraints. 
Therefore, the expected value of $L$ is at least $\sqrt{k/q}\cdot (1-o(1))q/16\geq 0.06\sqrt{kq}$, as claimed. 
\end{proof}

\ifx\mainfile\undefined

\end{document}

\else
\fi

\begin{proof}[Proof of Lemma~\ref{lem_2DORC_cc_lb}]
Assume there is a $\ANC$ protocol $2^{-0.2q\log\log n}$-solves $G_{\ORC}(k,q,n)$ with communication cost $(0.5\sqrt{kq}, 0.005q\log n, 0.0005q\log n)$. By Lemma~\ref{lem_es_lb} and Lemma~\ref{lem_ws_lb}, the probability that in a random input pair $(x, y)$ sampled from $\mu$, $x$ is not \typ{} or $y$ is not \ws{} is at most $1-2^{-0.4q\log\log n}-2^{-q\log^{500} n}$. By Lemma~\ref{lem_cm_rect}, for large enough $n$, $M(G_{\ORC})$ has a rectangle $R=X\times Y$ such that
\begin{enumerate}
\item
	every $x\in X$ is \typ{} and every $y\in Y$ is \ws{};
\item
	$R$ is $2^{-0.0006q\log n}$-\cm;
\item
	$\mu_x(X)\geq 2^{-0.6\sqrt{kq}}$;
\item
	$\mu_y(Y)\geq 2^{-0.006q\log n}$.
\end{enumerate}

However, by Lemma~\ref{lem_lb_cm_rect}, such rectangle cannot exist in $M(G_{\ORC})$. We have a contradiction. 
\end{proof}
}

\def\lemLowerProbeCompact{
\begin{lemma}\label{lem_lower_probe}
Let $\cO$ be a random sequence of operations sampled from $\cD$, let $I_A$ and $I_B$ be two consecutive intervals in $\cO$, 
such that $|I_B|\geq \sqrt{n}$ and $|I_A|\sim |I_B|\log^{c+1000}n$, and denote by $\cO_{<I_A}$ the sequence of operations preceding 
$I_A$. Then \emph{conditioned on} $\cO_{<I_A}$, the probability that all of the following events occur simultaneously is at most $2\cdot 2^{-0.2|I_B|\log\log n}$: 

\begin{tabular}{p{0.4\textwidth}p{0.5\textwidth}}
1) $|P(I_A)|\leq |I_A|\log^2 n$; & 3) $|P(I_A)\cap P(I_B)|\leq |I_B|\frac{\log n}{200(c+1002)\log\log n}$; \\
2)	$|P(I_B)|\leq |I_B|\log^2 n$; & 4) $D$ answers all queries in $I_B$ correctly. \\
\end{tabular}
\end{lemma}
}

\def\contLemLowerProbe{
Let $\cO$ be a random sequence of operations sampled from $\cD$, let $I_A$ and $I_B$ be two consecutive intervals in $\cO$, 
such that $|I_B|\geq \sqrt{n}$ and $|I_A|\sim |I_B|\log^{c+1000}n$, and denote by $\cO_{<I_A}$ the sequence of operations preceding 
$I_A$. Then \emph{conditioned on} $\cO_{<I_A}$, the probability that all of the following events occur simultaneously
\begin{enumerate}
\item
	$|P(I_A)|\leq |I_A|\log^2 n$,
\item
	$|P(I_B)|\leq |I_B|\log^2 n$, 
\item
	$|P(I_A)\cap P(I_B)|\leq |I_B|\frac{\log n}{200(c+1002)\log\log n}$, and
\item
	$D$ answers all queries in $I_B$ correctly
\end{enumerate}
is at most $2\cdot 2^{-0.2|I_B|\log\log n}$.
}

\def\lemLowerProbeProof{
\begin{proof}
Let $p$ denote the probability that all four events above occur. Consider the protocol $P_D$ on intervals $I_A$ and $I_B$, solving $G_{\ORC}(k,q,n)$ for $k\sim |I_A|\cdot\log^{-c} n$ and $q\sim |I_B|$.
Note that when the input pairs for $G_{\ORC}(k,q,n)$ are sampled uniformly at 
random (i.e., according to $\mu$), the corresponding operations $I_A$ and $I_B$ in the simulation $P_D$ are distributed according to $\cD$.
By Lemma~\ref{lem_sim_orc}, the first three conditions in the statement imply that the protocol $P_D$ satisfies the following conditions:
\begin{enumerate}
\item
	Alice sends at most $\frac{4q\log^2 n}{200(c+1002)\log\log n}\ll 0.1\sqrt{kq}$ bits in expectation,
\item
	Bob sends $$|P(I_A)\cap P(I_B)|\cdot \log\frac{e|P(I_A)|}{|P(I_A)\cap P(I_B)|}\leq 0.005q\log n$$ bits 
	(since the function $f(u, v)=u\cdot\log\frac{v}{u}$ is increasing in both $u$ and $v$ when $v>eu$), 
\item
Merlin sends  $$|P(I_A)\cap P(I_B)|\cdot \log\frac{e|P(I_B)|}{|P(I_A)\cap P(I_B)|}+O\left(\log n\right)<0.0005q\log n$$ bits.
\end{enumerate}

Conditioned on the the three events above, with probability at least $1/2$, 
Alice sends no more than $0.2\sqrt{kq}$ bits. Thus, by Lemma~\ref{lem_2DORC_cc_lb}, $p/2\leq 2^{-0.2q\log\log n}$. This proves the lemma.
\end{proof}
}

\iftoggle{focs}{Combining the communication lower bound for $G_{\ORC}(k,q,n)$ with the simulation argument in Appendix~\ref{app_sim_ORC} gives us a lower bound 
on the efficiency and accuracy of $D$ in $I_A$ and $I_B$. The proof of the following lemma can be found in Appendix~\ref{app_lem_lower_probe}.

\lemLowerProbeCompact
}{
\subsection{Proof of Theorem~\ref{thm2d}}\label{subsec_dslower}

Combining the communication lower bound for $G_{\ORC}(k,q,n)$ (Lemma \ref{lem_2DORC_cc_lb}) with 
Lemma \ref{lem_sim_orc} (the simulation argument in Section~\ref{subsect_sim_ORC}) gives us a lower bound 
on the efficiency and accuracy of $D$ in $I_A$ and $I_B$.

\begin{lemma}\label{lem_lower_probe}
\contLemLowerProbe
\end{lemma}

\lemLowerProbeProof
}

Using the above lemma, we are finally ready to prove our data structure lower bound for $\ORC$ (Theorem \ref{thm2d}). Intuitively, if the data structure $D$ correctly answers \emph{all} 
queries under $\cD$, then Lemma~\ref{lem_lower_probe} is essentially saying either $|P(I_A)|$ or $|P(I_B)|$ is large, or during $I_B$, $D$ reads at least $\Omega(|I_B|\cdot \log n/c\log\log n)$ cells that are also probed in $I_A$. If the former happens too often, it is not hard to see that $D$ makes too many probes in total. Otherwise, ``on average'' for every operation in $I_B$, $D$ must read at least $\Omega(\log n/c\log\log n)$ cells whose last probe was in $I_A$. This argument holds as long as $|I_A|\sim|I_B|\cdot \log^{c+O(1)} n$ and $|I_B|\geq \sqrt{n}$. Thus, during an operation, ``on average'' $D$ has to read $\Omega(\log n/c\log\log n)$ cells whose last probe was anywhere 
between $\sqrt{n}$ and $\sqrt{n}\cdot \log^{c+O(1)} n$ operations ago, $\Omega(\log n/c\log\log n)$ cells whose last probe was between 
$\sqrt{n}\cdot \log^{c+O(1)} n$ and $\sqrt{n}\cdot \log^{2(c+O(1))} n$ operations ago, and so on. All these sets of cells are disjoint, and there are 
$\Omega(\log n/c\log \log n)$ such sets. This gives us that ``on average'', each operation has to probe $\Omega((\log n/c\log \log n)^2)$ cells in total. 
While Lemma~\ref{lem_lower_probe} does not account for the number of cells probed during any particular operation (but only the total number of probes), 
 summing up the lower bounds for relevant interval pairs, the above argument gives an \emph{amortized} lower bound assuming $D$ correctly answers all queries. 

When $D$ is allowed to err, a natural approach is to partition the sequence into disjoint intervals $\{I\}$, and interpret the overall success probability as the product of 
conditional success probabilities for queries in $I$ conditioned on the event that $D$ succeeds on all queries preceding $I$. When the overall success probability 
is ``non-trivial'', there will be a constant fraction of $I$'s with ``non-trivial'' success probability conditioned on succeeding 
on all previous intervals. As Lemma~\ref{lem_lower_probe} also holds for $D$ that is correct with exponentially in 
$|I_B|\log\log n$ small probability, the argument outlined in the last paragraph still goes through. A more careful argument proves the theorem.
\def\thmtwoDProof{
\begin{proof}[Proof of Theorem~\ref{thm2d}]
We shall decompose the operation sequence into many (possibly overlapping) consecutive intervals, and then 
apply Lemma~\ref{lem_lower_probe} to obtain a lower bound on the operational time (in terms of probes) the data structure spends 
on each interval. Summing the lower bounds together will yield the desired lower bound on the total number of probes. 
To this end, 
for $\gamma=\log^{c+1000}n$, consider the following decomposition:
\begin{itemize}
	\item
		For any (consecutive) interval of operations $I$, define $\dec(I):=(I_A,I_B)$, where $I_A$ is the first $|I|-\lceil|I|/\gamma\rceil$ operations in $I$ and $I_B$ is the last $\lceil|I|/\gamma\rceil$ operations in $I$. We have $I_A\cap I_B=\emptyset$, $I_A\cup I_B=I$, and $|I_A|\sim|I_B|\cdot \gamma$ for any large 
		enough $|I|$. 
	\item
		Let $\cI_0=\{\dec(\cO)\}$ be the singleton set $(I_A, I_B)$, where $I_A$ is the first $n-\lceil n/\gamma\rceil$ operations and 
		$I_B$ is the last $\lceil n/\gamma\rceil$ operations in a random sequence $\cO$ sampled from $\cD$. 
	\item
		For $i>0$, let us recursively define 
		$$\cI_i=\{\dec(I)=(I_A, I_B) : |I_B|\geq\sqrt{n} \text{ \; s.t\; }\exists I', (I, I')\in \cI_{i-1}\vee (I', I)\in \cI_{i-1}\}$$ 
		to be the decomposition into (disjoint) intervals obtained by ``refining" the decomposition $\cI_{i-1}$ (as in the following illustration).
\end{itemize}

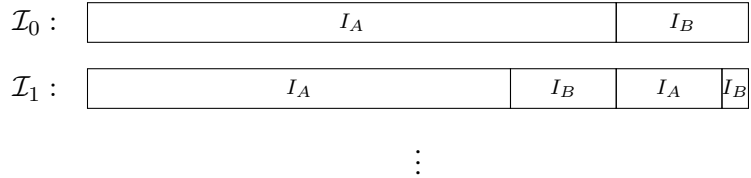
\begin{center}
\begin{tikzpicture}
	\node at (-20pt, 0pt) {$\cI_0:$};
	\node [draw, minimum width=200pt, inner sep=0pt, minimum height=15pt] at (100pt, 0pt) {\scriptsize $I_A$};
	\node [draw, minimum width=50pt, inner sep=0pt, minimum height=15pt] at (225pt, 0pt) {\scriptsize $I_B$};
	\node at (-20pt, -25pt) {$\cI_1:$};
	\node [draw, minimum width=160pt, inner sep=0pt, minimum height=15pt] at (80pt, -25pt) {\scriptsize $I_A$};
	\node [draw, minimum width=40pt, inner sep=0pt, minimum height=15pt] at (180pt, -25pt) {\scriptsize $I_B$};
	\node [draw, minimum width=40pt, inner sep=0pt, minimum height=15pt] at (220pt, -25pt) {\scriptsize $I_A$};
	\node [draw, minimum width=10pt, inner sep=0pt, minimum height=15pt] at (245pt, -25pt) {\scriptsize $I_B$};
	\node at (125pt, -50pt) {$\vdots$};
\end{tikzpicture}
\end{center}

The following claim, whose proof is deferred to Appendix~\ref{sec_proof_of_claim_game_sum}, 
states that for small enough $i$'s, 
the total number of operations in  ``Bob's intervals" ($I_B$) when summing up over all interval pairs in the set $\cI_i$, is large:

\begin{claim}\label{claim_game_sum}
For $i\leq 0.1\gamma\log n/\log\gamma$, we have
\[
	\sum_{(I_A, I_B)\in \cI_i}|I_B|\geq \frac{n}{2\gamma}.
\]
\end{claim}

Now, let $D$ be any dynamic data structure for $\ORC$, and 
define $\cE(I_A, I_B)$ to be the event that all four conditions of  Lemma~\ref{lem_lower_probe} occur with respect to 
a specific interval pair $I_A, I_B$ and the data structure $D$, namely:
\begin{enumerate}
\item
	$|P(I_A)|\leq |I_A|\log^2 n$,
\item
	$|P(I_B)|\leq |I_B|\log^2 n$, 
\item
	$|P(I_A)\cap P(I_B)|\leq |I_B|\frac{\log n}{200(c+1002)\log\log n}$, and
\item
	$D$ answers all queries in $I_B$ correctly.
\end{enumerate}

Consider the event $\cE_i$ that all queries are answered correctly and ``most" of $(I_A,I_B)\in \cI_i$ are efficient:
\begin{enumerate}
\item
	$\sum_{(I_A,I_B)\in \cI_i}|P(I_A)|\leq \frac{n\log^2 n}{8}$,
\item
	$\sum_{(I_A,I_B)\in \cI_i}|P(I_B)|\leq \frac{n\log^2 n}{8\gamma}$, 
\item
	$\sum_{(I_A,I_B)\in \cI_i}|P(I_A)\cap P(I_B)|\leq \frac{n\log n}{1600(c+1002)\gamma\log\log n}$, and
\item
	$D$ answers all queries in the sequence correctly.
\end{enumerate}

By Markov's inequality, $\cE_i$ implies that
\[
\begin{aligned}
	\sum_{\stackrel{(I_A,I_B)\in \cI_i:}{|P(I_A)|>|I_A|\log^2 n}}|I_B|&\leq \frac{1}{\gamma\log^2 n}\sum_{\stackrel{(I_A,I_B)\in \cI_i:}{|P(I_A)|>|I_A|\log^2 n}}|P(I_A)| 
	\leq \frac{1}{\gamma\log^2 n}\cdot \frac{n\log^2 n}{8} 
	=\frac{n}{8\gamma}.
\end{aligned}
\]
Similarly, we have
\[
	\sum_{\stackrel{(I_A,I_B)\in \cI_i:}{|P(I_B)|>|I_B|\log^2 n}}|I_B|\leq \frac{n}{8\gamma}
\text{\;\;\;\;\;\;\;\;\;\;\;\;\;\;\;\; and \;\;\;\;}
	\sum_{\stackrel{(I_A,I_B)\in \cI_i:}{|P(I_A)\cap P(I_B)|>|I_B|\frac{\log n}{200(c+1002)\log\log n}}}|I_B|\leq \frac{n}{8\gamma}.
\]
Therefore, by Claim~\ref{claim_game_sum} and a union bound, the event $\cE_i$ implies the event 
``$\sum_{(I_A,I_B)\in \cI_i:\cE(I_A,I_B)}|I_B|\geq \frac{n}{8\gamma}$" whenever $i\leq 0.1\gamma\log n/\log\gamma$. 

We now want to use this fact together with Lemma~\ref{lem_lower_probe} to conclude that $\cE_i$ cannot occur too often. 
Indeed, Lemma~\ref{lem_lower_probe} asserts that the event $\cE(I_A, I_B)$ occurs with extremely low probability,
even conditioned on all operations before $I_A$. 
Since all the intervals in $\cI_i$ are disjoint by construction, this gives us an upper bound on the probability of $\cE_i$ : 
\[
\begin{aligned}
\Pr[\cE_i]&\leq \Pr\left[\sum_{(I_A,I_B)\in \cI_i:\cE(I_A,I_B)}|I_B|\geq \frac{n}{8\gamma}\right] \\
&= \Pr\left[\exists S\subseteq\cI_i, \sum_{(I_A, I_B)\in S}|I_B|\geq \frac{n}{8\gamma}, \forall (I_A, I_B)\in S, \cE(I_A, I_B)\right] \\
&\leq \sum_{\stackrel{S\subseteq\cI_i,}{\sum_{(I_A, I_B)\in S}|I_B|\geq \frac{n}{8\gamma}}}\Pr\left[\bigwedge_{(I_A, I_B)\in S} \cE(I_A, I_B)\right] && \textrm{(by union bound)}\\
&= \sum_{\stackrel{S\subseteq\cI_i,}{\sum_{(I_A, I_B)\in S}|I_B|\geq \frac{n}{8\gamma}}}\prod_{(I_A,I_B)\in S}\Pr\left[\cE(I_A, I_B)\,\,\,\middle|\,\,\,\, \bigwedge_{\stackrel{(I_A', I_B')\in S}{(I_A', I_B')\textrm{ before }(I_A, I_B)}} \cE(I_A', I_B')\right] \\
&\leq \sum_{\stackrel{S\subseteq\cI_i,}{\sum_{(I_A, I_B)\in S}|I_B|\geq \frac{n}{8\gamma}}}\prod_{(I_A,I_B)\in S}\left(2\cdot 2^{-0.2|I_B|\log\log n}\right) && \textrm{(by Lemma~\ref{lem_lower_probe})}\\
&\leq \sum_{\stackrel{S\subseteq\cI_i,}{\sum_{(I_A, I_B)\in S}|I_B|\geq \frac{n}{8\gamma}}}2^{|S|}2^{-0.025\frac{n}{\gamma}\log\log n} \\
&\leq 4^{|\cI_i|}2^{-0.025\frac{n}{\gamma}\log\log n}. 
\end{aligned}
\]

Since all $I_B$'s in $\cI_i$ are disjoint and have length at least $\sqrt{n}$, $|\cI_i|\leq \sqrt{n}$. Thus, for $n$ large enough, $\Pr[\cE_i]\leq 2^{-0.02\frac{n}{\gamma}\log\log n}$.

Finally, recall that $\cE_i$ is the event that all queries are answered correctly and most interval pairs in $\cI_i$ are efficient. 
To finish the proof, we claim that the efficiency 
of interval pairs in all $\cI_i$ characterizes the overall efficiency:

\begin{claim}\label{cl_E_i_ocurs}
If $D$ probes $o(n(\log n/c\log\log n)^2)$ cells and is correct on all queries, then for some $i\leq 0.1\gamma\log n/\log\gamma$, 
$\cE_i$ occurs. 
\end{claim}

Indeed, consider the contrapositive statement that non of $\cE_i$ occurs. Then at least one of the following events must occur :
\begin{enumerate}
\item
	Some query is not answered correctly.
\item
	For some $i$, $\sum_{(I_A,I_B)\in \cI_i}|P(I_A)|>\frac{n\log^2 n}{8}$. Since all $I_A$'s are disjoint, this already implies that $D$ probes at too many cells.
\item
	There are least $0.05\gamma\log n/\log\gamma$ different $i$'s for which $\sum_{(I_A,I_B)\in \cI_i}|P(I_B)|> \frac{n\log^2 n}{8\gamma}$. 
	Since each operation can only appear in $\leq \log n/\log\gamma$ different $I_B$'s, the total number of probes is at least \[
		\frac{\log\gamma}{\log n}\cdot (0.05\gamma\log n/\log\gamma)\cdot \frac{n\log^2 n}{8\gamma}\geq \Omega(n\log^2 n).
	\]
\item
	There are least $0.05\gamma\log n/\log\gamma$ different $i$'s for which  
	$$\sum_{(I_A,I_B)\in \cI_i}|P(I_A)\cap P(I_B)|>\frac{n\log n}{200(c+1002)\gamma\log\log n}.$$ 
	By construction of the $\cI_i$'s, each probe will appear only once across all interval pairs. The total number of probes is at least 
	\[
		0.05\gamma\log n/\log\gamma\cdot \frac{n\log n}{200(c+1002)\gamma\log\log n}\geq \Omega\left(n\left(\frac{\log n}{c\log\log n}\right)^2\right).
	\]
\end{enumerate}

The above asserts that either $D$ is wrong on some query or $D$ probes $\Omega(n(\log n/c\log\log n)^2)$ cells, which finishes 
the proof of Claim \ref{cl_E_i_ocurs}.  

By a union bound over all $\cE_i$'s, we conclude that the probability that $D$ probes $o(n(\log n/c\log\log n)^2)$ cells and is correct on all queries is 
at most $$\sum_{i<0.1\gamma\log n/\log\gamma}\Pr[\cE_i]<2^{-n/\log^{c+O(1)} n},$$ which completes the proof of the entire theorem.
\end{proof}
}%
\iftoggle{focs}{The formal proof can be found in Appendix~\ref{app_thm2d_proof}.}
{We now turn to formalize the intuition above, by showing how to combine the arguments in the last two paragraphs.

\thmtwoDProof
}

\bibliographystyle{alpha}
\bibliography{refs}

\appendix


\iftoggle{focs}{

}


\section{Proof of Proposition~\ref{prop_3dorc}} \label{sec_append_prop_3dorc}

\newcommand{\tstat}{t_{\mathrm{stat}}}
\newcommand{\Dstat}{D_{\mathrm{stat}}}

Let $\Dstat$ be a (zero-error) data structure for the \emph{static} $\TDORC$ problem, that uses $s(m)$ memory cells and $\tstat(m)$ 
probes to answer any query on $m$ points. To solve the dynamic $\ORC$ problem on a sequence of $n/\log^c n$ updates and $n-n/\log^c n$ queries, we \emph{guess} all updates before the first operation. Let the $i$-th update we guessed be \texttt{update($r_i$, $c_i$, $w_i$)}. Then we use $\Dstat$ to build a static data structure on points $(i,r_i,c_i)$ with weight $w_i$, and write to the memory. This step costs $s(n/\log^c n)$ probes. Note that the cell-probe model does not charge for the actual construction time of the data structure, but only for the number of probes to the memory. In addition, we also maintain a counter $q$ in the memory, recording the number of updates performed so far, which is initialized to be $0$. 

To do an update, we simply increment the counter $q$ by one. To answer a query \texttt{query($r$, $c$)}, we first read the value of $q$ from the memory, then query $\Dstat$ the sum of weights of points dominated by $(q, r, c)$. This corresponds to asking $\Dstat$ what is the sum of weights of points appeared in the first $q$ updates (we guessed) that are dominated by $(r, c)$, which costs $\tstat(n/\log^c n)$ probes. In total, we spend $s(n/\log^c n)+n/\log^c n+n\cdot\tstat(n/\log^c n)$ probes.

If we happen to guess all $n/\log^c n$ update correctly, then all $n-n/\log^c n$ queries will be answered correctly, which happens with probability $n^{-3n/\log^c n}$. 
This is an upper bound on the probability asserted by the strengthened version of Theorem~\ref{thm2d}, that all queries are answered correctly and $o(n(\log n/c\log\log n)^2)$ cells are probed. In particular, we then must have $$s(n/\log^c n)+n\cdot\tstat(n/\log^c n)\geq \Omega(n(\log n/c\log\log n)^2).$$
Thus, if $s(m)\leq O(m\log^c m)$, we must have $\tstat(m)\geq \Omega((\log m/c\log\log m)^2)$. 

\subsection{Challenges in improving Theorem~\ref{thm2d}}\label{discussion}
Although only a slightly strengthened version of Theorem~\ref{thm2d} is required by Proposition~\ref{prop_3dorc}, making the improvement still seems non-trivial, as it forces us to \emph{break} Yao's Minimax Principle in the cell-probe model. In the cell-probe model, one direction of the principle is still true: A lower bound for deterministic data structures on a fixed hard input distribution is always a lower bound for any randomized data structure on its worst-case input. This is the direction that we (and also many previous works) use in the proof. But the other direction may no longer hold. One way to see it is that when we try to fix the random bits used by a randomized data structure, its running time may drop significantly. If the random bits are fixed, they can be hard-wired to the data structure, and can be accessed during the operations for free. On the other hand, if they are generated on the fly during previous operations, the data structure has to probe memory cells to recover them, because it does not remember anything across the operations by definition. A randomized data structure is a convex combination of deterministic data structures. Thus, in the cell-probe model, the function that maps a data structure to its performance on a fixed input may not be linear, i.e., the performance of a convex combination of deterministic data structures can be strictly larger than the same convex combination of the performances of deterministic data structures on a fixed input. In contrast,  the proof of Yao's Minimax Principle relies on 
the linearity of this function.

Indeed, for the dynamic $\ORC$ problem and any distribution over operation sequences with $n/\log^c n$ updates and $n-n/\log^c n$ queries, we can solve it trivially if only $n^{-3n/\log^c n}$ correct probability is required: hard-wire the most-likely sequence of $n/\log^c n$ updates, and answer all queries based on it. Thus, each update and query can be done in constant time. When the most-likely sequence of updates occurs (with probability $\geq n^{-3n/\log^c n}$), all queries will be answered correctly. Thus, to improve Theorem~\ref{thm2d},  one would have to design 
a ``data-structure-dependent" hard distribution, adversarially tailored to each data structure we are analyzing, and carrying out such argument seems to require new ideas. 

\iftoggle{focs}{
\section{Proof of Lemma~\ref{lem_sim_general}}\label{app_lem_sim_general}
\lemSimGeneral

\section{Efficient $\ANC$ Protocols Induce Large Biased Rectangles}\label{app_lem_monochromatic_rect} 
\restateMonochromaticRect
\lemMonochromaticRectFull

\section{Efficient Data Structure Simulation in the $\ANC$ Model}\label{app_sim_ORC}
\subsectSimORC

\section{$\ANC$ Communication Complexity of Orthogonal Range Counting}\label{app_comm_lower}
\begin{restate}[Lemma~\ref{lem_2DORC_cc_lb}]
\contlemTwoDORCcclb
\end{restate}

The obvious strategy for proving this lemma is to use the argument in Section~\ref{subsect_cm_rect}, which asserts that an efficient $\ANC$ protocol implies a large \cm{} 
rectangle. Therefore, ruling out the existence of a large \cm{} rectangle in $M(G_{\ORC})$ would give us a communication lower bound on $G_{\ORC}$.  

Unfortunately, $G_{\ORC}(k,q,n)$ does in fact contain large \cm{} rectangles. For example, when all of Bob's $q$ points have $r$-coordinate smaller than $n/\log^{\Theta(1)} n$, Alice does not have to tell Bob any information about her points with $r$-coordinate greater than that quantity. Thus, in expectation, Alice only needs to speak $k/\log^{\Theta(1)} n$ bits, and this case happens with $2^{-\Theta(q\log\log n)}$ probability over a random Bob's input. In the other word, there is a \cm{} rectangle of size $2^{-k/\log^{\Theta(1)} n}\times 2^{-\Theta(q\log\log n)}$. We cannot hope to prove a communication lower bound higher than $(k/\log^{\Theta(1)} n, \Theta(q\log\log n), 0)$ using this approach alone.

\begin{proof}[Proof of Lemma~\ref{lem_2DORC_cc_lb}]
Assume there is a $\ANC$ protocol $2^{-0.2q\log\log n}$-solves $G_{\ORC}(k,q,n)$ with communication cost $(0.5\sqrt{kq}, 0.005q\log n, 0.0005q\log n)$. By Lemma~\ref{lem_es_lb} and Lemma~\ref{lem_ws_lb}, the probability that in a random input pair $(x, y)$ sampled from $\mu$, $x$ is not \typ{} or $y$ is not \ws{} is at most $1-2^{-0.4q\log\log n}-2^{-q\log^{500} n}$. By Lemma~\ref{lem_cm_rect}, for large enough $n$, $M(G_{\ORC})$ has a rectangle $R=X\times Y$ such that
\begin{enumerate}
\item
	every $x\in X$ is \typ{} and every $y\in Y$ is \ws{};
\item
	$R$ is $2^{-0.0006q\log n}$-\cm;
\item
	$\mu_x(X)\geq 2^{-0.6\sqrt{kq}}$;
\item
	$\mu_y(Y)\geq 2^{-0.006q\log n}$.
\end{enumerate}

However, by Lemma~\ref{lem_lb_cm_rect}, such rectangle cannot exist in $M(G_{\ORC})$. We have a contradiction. 
\end{proof}

\section{Proof of Lemma~\ref{lem_lower_probe}}\label{app_lem_lower_probe}
\begin{restate}[Lemma~\ref{lem_lower_probe}]
\contLemLowerProbe
\end{restate}
\lemLowerProbeProof

\section{Proof of Theorem~\ref{thm2d}}\label{app_thm2d_proof}
\begin{restate}[Theorem~\ref{thm2d}]
\thmtwodStatement
\end{restate}

\thmtwoDProof
}


\section{Proof of Claim \ref{claim_game_sum}} \label{sec_proof_of_claim_game_sum}

\begin{proof}[Proof of Claim~\ref{claim_game_sum}]
The procedure of generating the sets $\cI_i$ can be modelled as a binary tree. The root of the tree is $(I_A, I_B)=\dec(\cO)$. Its left child is $\dec(I_A)$, and its right child is $\dec(I_B)$. In general, for each node $(I, I')$, its left child is $\dec(I)$ and its right child is $\dec(I')$. We keep expanding until either $|I'|<\sqrt{n}$ or the node is in depth $i$. It is easy to verify that
\begin{itemize}
\item
	the sum of $|I_B|$ over all leaves $(I_A, I_B)$ is $\sim n/\gamma$,
\item
	$\sum_{(I_A,I_B)\in\cI_i}|I_B|$ is just the sum of $|I_B|$ over all leaves with $|I_B|\geq\sqrt{n}$. 
\end{itemize}

Thus, it is sufficient to bound the number of leaves $(I_A, I_B)$ with $|I_B|<\sqrt{n}$. Fix a leaf, consider the path from root to it. Every time the path follows a left-child-edge, $|I_B|$ shrinks by a factor of $1-1/\gamma$. Every time it follows a right-child-edge, $|I_B|$ shrinks by a factor of $1/\gamma$. Since we stop expanding the tree as soon as $|I_B|<\sqrt{n}$, the path can follow a right-child-edge at most $\lceil \log (\sqrt{n}/\gamma)/\log\gamma\rceil$ times (and stops as soon as it follows the $\lceil \log (\sqrt{n}/\gamma)/\log\gamma\rceil$ one). Thus, there are at most 
\[
\begin{aligned}
{\leq 0.1\gamma\log n/\log\gamma\choose \leq \lceil \log (\sqrt{n}/\gamma)/\log\gamma\rceil-1}&\leq {\leq 0.1\gamma\log n/\log\gamma\choose \leq \log (\sqrt{n}/\gamma)/\log\gamma} \\
&\leq (0.2e\gamma)^{\log (\sqrt{n}/\gamma)/\log\gamma} \\
&< 0.6^{\log (\sqrt{n}/\gamma)/\log\gamma}\cdot\frac{\sqrt{n}}{\gamma} \ll \frac{\sqrt{n}}{2\gamma}
\end{aligned}
\]
leaves in total. By above observation, $\sum_{(I_A,I_B)\in\cI_i}|I_B|\geq n/\gamma-\sqrt{n}\cdot \sqrt{n}/2\gamma\geq n/2\gamma$. 
\end{proof}

\end{document}